\documentclass[journal]{IEEEtran}
\usepackage{cite}
\ifCLASSINFOpdf
\else
\fi

\usepackage[T1]{fontenc}
\usepackage[utf8]{inputenc}
\usepackage{fancybox}
\usepackage{tikz}
\usetikzlibrary{shapes,arrows.meta,shadows}
\usepackage{graphicx}

\usepackage{amsmath,bm,times}
\usepackage{paralist}
\usepackage{amsfonts}
\usepackage{amssymb}
\usepackage{array}
\usepackage{amsthm}
\usepackage{epstopdf}

\usepackage{multirow}
\theoremstyle{definition}
\newtheorem*{definition*}{Definition}
\newtheorem{theorem}{Theorem}
\newtheorem{result}{Result}
\newtheorem{proposition}{Proposition}
\newtheorem{corollary}{Corollary}[proposition]

\DeclareMathOperator*{\argmin}{arg\,min}
\usepackage{algorithmicx}
\usepackage{algorithm}
\usepackage{textcomp}
\def\BibTeX{{\rm B\kern-.05em{\sc i\kern-.025em b}\kern-.08em
		T\kern-.1667em\lower.7ex\hbox{E}\kern-.125emX}}
\usepackage{paralist}
\newcommand{\oblvl}{O}

\newcommand{\Root}{x}
\newcommand{\Obs}{y}
\newcommand{\prob}{\mathbb{P}}
\newcommand{\state}{x}
\newcommand{\Time}{k}

\newcommand{\statespace}{\mathcal{X}}
\newcommand{\tp}{P}
\newcommand{\belief}{\pi}
\newcommand{\Belief}{\Pi}
\newcommand{\action}{u}

\newcommand{\obp}{B}
\newcommand{\filter}{T}

\newcommand{\E}{\mathbb{E}}
\newcommand{\obs}{y}
\newcommand{\filternorm}{\sigma}

\newcommand{\policy}{\mu}

\newcommand{\actionspace}{\mathcal{U}}
\newcommand{\actiondim}{N}
\newcommand{\cost}{C}
\newcommand{\costm}{C}

\newcommand{\errorh}{\eta}
\newcommand{\weight}{w}

\newcommand{\discount}{\rho}
\newcommand{\discountedcost}{J}
\newcommand{\optpolicy}{\policy^*}
\newcommand{\valuefunction}{V}
\newcommand{\lbpolicy}{\bar{\policy}}
\newcommand{\Beliefset}{\Belief^s}

\newcommand{\epserr}{\mathcal{L}}

\newcommand{\pmfnew}{\Lambda}

\newcommand{\nnode}{\mathcal{N}}

\usepackage{accents}
\usepackage{environ}
\usepackage{comment}
\makeatletter
\newsavebox{\measure@tikzpicture}
\NewEnviron{scaletikzpicturetowidth}[1]{%
	\def\tikz@width{#1}%
	\def\tikzscale{1}\begin{lrbox}{\measure@tikzpicture}%
		\BODY
	\end{lrbox}%
	\pgfmathparse{#1/\wd\measure@tikzpicture}%
	\edef\tikzscale{\pgfmathresult}%
	\BODY
}
\makeatother
\tikzset{>={Latex[width=3mm,length=3mm]}}
\pgfdeclarelayer{background}
\pgfdeclarelayer{foreground}
\pgfsetlayers{background,main,foreground}

\tikzstyle{sensor}=[draw, fill=blue!20, text width=5em, 
text centered, minimum height=2.5em,drop shadow]
\tikzstyle{ann} = [above, text width=5em, text centered]
\tikzstyle{wa} = [sensor, text width=6.8em, fill=red!20, 
minimum height=3em, rounded corners, drop shadow]
\tikzstyle{sc} = [sensor, text width=13em, fill=red!20, 
minimum height=10em, rounded corners, drop shadow]

\ifCLASSOPTIONcompsoc
  \usepackage[nocompress]{cite}
\else
  \usepackage{cite}
\fi

\begin{document}
\title{Adaptive Polling in Hierarchical Social Networks using Blackwell Dominance}
\author{{\hspace{2cm}}Sujay Bhatt and Vikram Krishnamurthy,~\IEEEmembership{Fellow,~IEEE}
	\IEEEcompsocitemizethanks{\IEEEcompsocthanksitem S. Bhatt  and V. Krishnamurthy are with the Department of Electrical and Computer Engineering, Cornell Tech, Cornell University, NY, NY 10044.\protect\\
		E-mail: sh2376@cornell.edu,~vikramk@cornell.edu
		\IEEEcompsocthanksitem This research was supported, in part by,  the U. S. Army Research Office under grant 12346080, National Science Foundation under grant 1714180, and Air Force Office of Scientific Researcher under grant FA9550-18-1-0007.}}
\maketitle
\begin{abstract}
Consider a population of individuals that observe an underlying state of nature that evolves over time. The population is classified into different levels depending on the hierarchical influence that dictates how the individuals at each level form an opinion on the state. The population is sampled sequentially by a pollster and the nodes (or individuals) respond to the questions asked by the pollster. This paper considers the following problem-- \textit{How should the pollster poll the hierarchical social network to estimate the state while minimizing the polling cost (measurement cost and uncertainty in the Bayesian state estimate)?} 

This paper proposes adaptive versions of the following polling methods --Intent Polling, Expectation Polling, and the recently proposed Neighbourhood Expectation Polling -- to account for the time varying state of nature and the hierarchical influence in social networks. 
The adaptive polling problem in a hierarchical social network is formulated as a partially observed Markov decision process (POMDP). 
Our main results exploit the structure of the polling problem, and determine novel conditions for Blackwell dominance to construct myopic policies that provably upper bound the optimal policy
of the adaptive polling POMDP. The LeCam deficiency is used to determine approximate Blackwell dominance for general polling problems. These Blackwell dominance conditions also facilitate the comparison of R{\'e}nyi Divergence and Shannon capacity of more general channel structures that arise in hierarchical social networks.
Numerical examples are provided to illustrate the adaptive polling policies with parameters estimated from YouTube data.
\end{abstract}

\begin{IEEEkeywords}
Adaptive polling, POMDP, structural result, Blackwell dominance, myopic policy, intent polling, expectation polling, Shannon capacity
\end{IEEEkeywords}



\section{Introduction}\label{sec:introduction}

Blackwell dominance and LeCam deficiency are widely used in statistical analysis of estimators \cite{Lec12,Lec12a}, in characterizing correlated and Nash equilibria in games \cite{Ber16}, and in stochastic control \cite{Rei91,Kri16}. Blackwell dominance also has deeper information theoretic interpretations \cite{CKZ98}. In this paper, we  use Blackwell dominance to construct efficient polling strategies for hierarchical social networks. The Blackwell dominance conditions we construct for adaptive polling also have simple information theoretic interpretations in terms of capacities of more general channel structures.\\
Polling has numerous applications such as predicting the outcome of an election, estimating the fraction of supporters of a particular party or fraction that believe in climate change, and predicting the success of a particular product. However, the existing polling methods do not consider the effect of influence that is inherently present in social networks. Many social networks have a hierarchical structure; see for example~\cite{LMS02, ALS03, Shi10, BJA11, Boy15}. This hierarchical structure specifies the influence exerted by the higher levels on the lower levels. This influence alters the opinions of the lower level nodes and hence affects the prediction or the poll estimate. \\
This paper takes the hierarchical structure into account to devise adaptive (feedback control based) polling strategies in a hierarchical social network to  minimize the polling cost (measurement cost and uncertainty in the Bayesian state estimate). We call such feedback control based polling as  \textit{Adaptive Polling}; see Fig.\ref{fig:HSN}.  
The adaptive polling problem is formulated as a Partially Observed Markov Decision Process (POMDP). POMDPs provide a principled mathematical framework to deal with sequential decision making problems with feedback control in partially observed domains. The actions taken by the pollster can influence the underlying state, noisy state-observations or both. The goal of the pollster is to choose an action, based on the history of past actions and observations, that minimizes the expected costs incurred over time.

\subsection{Context. Blackwell Dominance}
In general,  POMDPs are computationally intractable{\footnote{They are PSPACE hard requiring exponential computational cost (in sample path length) and memory \cite{Lov91a,Kri16}.}} to solve~\cite{PT87}. The main contribution of this paper is to exploit the structure of the social influence network to construct computationally efficient \textit{myopic policies} that provably upper bound the optimal polling policy.  Construction of such myopic bounds involves using the concept of Blackwell dominance of the observation
likelihoods. 

Since the main results of the paper rely on Blackwell dominance, for convenience, we now define Blackwell dominance and its important information theoretic consequences. 
Blackwell dominance formalizes the notion of which distribution (stochastic matrix) is \textit{more informative} than the other. A stochastic\footnote{A $\mathcal{X} \times \mathcal{Y}$ matrix $B$ is (row) stochastic if $\sum_j B_{ij} = 1$ for all $i \in \mathcal{X}$, $j \in \mathcal{Y}$, and $B_{ij} \in [0,1]$.}  matrix $B(1) \in \mathbb{P}( \mathcal{Y}^{(1)} |  \mathcal{X})$ Blackwell dominates~\cite{Bla53, Kri16}  another stochastic matrix $B(2) \in \mathbb{P}( \mathcal{Y}^{(2)} |  \mathcal{X})$ written as~$B(1) \succeq_B B(2)$ (or $B(1)$ is more informative than $B(2)$), if 
\begin{equation} \label{eq:PoPr}
B(2)= B(1)R ,~\text{for any stochastic matrix}~R.
\end{equation}

Blackwell dominance also has an information theoretic consequence: Consider the textbook Discrete Memoryless Channel (DMC) \cite{CT02} with input alphabet $\mathcal{X}$ and output alphabet $\mathcal{Y}$ denoted as $\mathbb{P}( \mathcal{Y} |  \mathcal{X})$. Let $I(\mathcal{X};\mathcal{Y})$ denote the mutual information of the DMC. 
The post-processing of channel $B(1)$ in (\ref{eq:PoPr}) is written as $\mathcal{X} \rightarrow \mathcal{Y}^{(1)} \rightarrow \mathcal{Y}^{(2)}$. Then  from Data Processing Inequality~\cite{CT02}, it follows that
\begin{equation} \label{eq:SIq}
B(1) \succeq_B B(2) \Rightarrow I(\mathcal{X};\mathcal{Y}^{(1)}) \geq  I(\mathcal{X};\mathcal{Y}^{(2)}).
\end{equation}
Theorem~\ref{eq:SCEq} below provides a relation between Blackwell Dominance and Shannon capacity.
\begin{theorem}[\cite{CKZ98,Rag11,RBOJBW17}] \label{eq:SCEq}
	For any two conditional distributions $B(1) \in \mathbb{P}( \mathcal{Y}^{(1)} |  \mathcal{X})$ and $B(2) \in \mathbb{P}( \mathcal{Y}^{(2)} |  \mathcal{X})$,
	\begin{equation}
	B(1) \succeq_B B(2) \Rightarrow \mathcal{C}^{(1)} \geq \mathcal{C}^{(2)},
	\end{equation}
	where the Shannon capacity $\mathcal{C}^{(i)}$ of a DMC is defined as 
	\begin{equation} \label{eq:SCap}
	\mathcal{C}^{(i)} = \sup_{p_{\mathcal{X}}(x)} I(\mathcal{X};\mathcal{Y}^{(i)}),~~i = 1,2.
	\end{equation}
	Here $p_{\mathcal{X}}(x)$ is the marginal distribution over the input alphabet~$\mathcal{X}$.
\end{theorem}
In this paper, we will characterize the capacity for more general channel structures that arise in polling hierarchical social networks. Also, Blackwell dominance is used to order the R{\'e}nyi Divergence \cite{CT02} of the observation likelihoods of these channels. These information theoretic consequences 
provide a ranking of these general channel structures in the order of their ability to distinguish the states.

\subsection{Main Results and Organization}
(i)~In Sec.\ref{sec:CSHN}, the underlying state is modeled as a Markov chain and the adaptive polling problem is formulated as a POMDP.
	Open loop polling, where polling at a particular instant is not influenced by the information previously collected, is ineffective when the states evolve over time. In comparison, the  proposed adaptive (feedback) polling procedure utilizes information previously collected to poll at the next instant. In a hierarchical social network, the nodes at higher levels in the hierarchy are more influential and so provide more accurate information on the underlying state than the lower levels(see Fig.\ref{fig:HSN}).
	The  proposed adaptive polling mechanism for hierarchical social networks also takes this into account. \\ We formulate adaptive generalizations of the \textit{Intent Polling} and \textit{Expectation Polling} methods{\footnote{
			{\em Intention}: Who will you vote for?\\
			{\em Expectation}: Who do you think will win?}} \cite{RW11} in Sec.\ref{sec:CIP} and Sec.\ref{sec:CEP}, and Neighbourhood Expectation Polling \cite{NV18}
	based on \textit{Friendship Paradox}\footnote{The friendship paradox \cite{Fel91}  refers to the phenomenon that on average
		your friends have more friends than you; see Sec.\ref{sec:CFP} for a more precise statement. Neighborhood Expectation Polling asks: What is a your estimate of the fraction of votes for a particular candidate?} in Sec.\ref{sec:CFP}. \\ 
	(ii)~\textit{Blackwell Dominance  in Hierarchical Networks}: 
	As mentioned above, in general, solving a POMDP is computationally intractable (see Footnote~$1$). A key property of our adaptive polling POMDP is that it exhibits a Blackwell dominance structure. For such POMDPs, a myopic policy provably forms an upper bound to the optimal policy
	(Theorem~\ref{thm:MT1}). For the three adaptive polling POMDPs considered, namely, intent polling, expectation polling and neighbourhood expectation polling, we present several novel
	sufficient conditions for Blackwell dominance involving matrix polynomial functions (Proposition~\ref{thm:CIP}) and ultrametric matrices (Proposition~\ref{thm:CEP}). This in turn facilitates the comparison of R{\'e}nyi Divergence and Shannon capacity of more general channels that arise naturally in hierarchical social networks. For example, Proposition~\ref{thm:CIP} provides an interesting link between Hurwitz (stable) polynomials and Shannon capacity. \\
(iii)~\textit{Approximate Blackwell Dominance}: Blackwell dominance induces  a partial order between two stochastic matrices implying not every pair is comparable. However, the upper bounds in Theorem~\ref{thm:MT1} provide sufficient motivation to find a pair of matrices that are close to the given pair and are Blackwell comparable. Sec.\ref{sec:APB} defines the notion of closeness between stochastic matrices using \textit{Le Cam deficiency}. Using this notion of approximate Blackwell dominance, we discuss how to design polling POMDPs that allow a comparison between the three proposed polling mechanisms. The performance bounds of the mis-specified POMDP model and policy are provided. We also discuss ordinal sensitivity in polling hierarchical networks, where we show that some networks are inherently more expensive to poll than others. \\
Sec.\ref{sec:NEx} provides numerical examples that illustrate the myopic polling policies.

\subsection{Related Literature}
\cite{KH14} analyzes a Bayesian approach to intent and expectation polling, but without feedback control. Polling has been considered in \cite{RW11} and a comparison of intent and expectation polling (non-Bayesian) algorithms is discussed{\footnote{\cite{RW11} analyzes all US presidential electoral college results from $1952-2008$ where both intention and expectation polling were conducted and shows a remarkable
		result: In $77$ cases where expectation and intent polling pointed to different winners, expectation polling was accurate $78\%$ of the time! }}. A trade-off between number of polled individuals and the bias introduced due to the network structure is discussed in \cite{DKS12}. Adaptive or feedback based approaches to similar problems have been considered in~\cite{Sat94,Lin13}. 

\section{Adaptive Polling in Hierarchical Social Networks} \label{sec:CSHN}
This section formulates the adaptive polling problem as a partially observed Markov decision process (POMDP). Sec.\ref{sec:MN} introduces the model for the adaptive polling problem and Sec.\ref{sec:PMDP} formulates the adaptive polling problem as a POMDP. Then the main result, namely, sufficient conditions on the model parameters that enable us to compute myopic upper bounds for the optimal policy, are provided using Blackwell dominance.
\begin{figure}[h!]
	\begin{scaletikzpicturetowidth}{0.4\textwidth}
		\hspace{-0.1cm}  \begin{tikzpicture}[scale=\tikzscale]
		\node (wa) [wa]  {Markov Process\\$P$};
		\path (wa.east)+(4.5,2) node (asr1) [sensor] {$\text{Level}~0$};
		\path (wa.east)+(4.5,0) node (asr2)[sensor] {$\text{Level}~1$};
		\path (wa.east)+(4.5,-1.5) node (dots)[ann] {$\vdots$}; 
		\path (wa.east)+(4.5,-2.5) node (asr3)[sensor] {$\text{Level}~N$};

		
		\path [draw, ->] (asr1.south) -- node [above] {} 
		(asr2.90) ;
		\path [draw, ->] (dots.south) -- node [above] {} 
		(asr3.90);
		(vote.west);

		\path (asr3.south) +(0,-0.4) node (asrs) {Hierarchical Social Network};
		\path (asrs.south) + (0,-3) node (vote) [sensor] {Filter \& \\ Controller};
		
		\begin{pgfonlayer}{background}
		\path (asrs.west |- asr1.north)+(-0,0.3) node (a) {};
		\path (wa.south -| wa.east)+(+0.5,-0.3) node (b) {};
		\path (asrs.east |- asrs.east)+(+0,-0.5) node (c) {};
		
		\path[fill=yellow!20,rounded corners, draw=black!50, dashed]
		(a) rectangle (c);           
		\path (asr1.north west)+(-0.2,0.2) node (a) {};
		
		\end{pgfonlayer}
		\path [draw, ->,very thick] (wa.east) -- node [below] {State~$(x_k)$}  ++(2.2,0);
		\path (vote.south) +(-0,-0.3) node (pol) {Pollster};
		\begin{pgfonlayer}{background}
		\path (vote.west |- vote.north)+(-0.5,0.3) node (m1) {};
		\path (vote.south -| vote.east)+(+0.5,-0.3) node (m2) {};
		\path (vote.east |- vote.east)+(+0.5,-1.1) node (m3) {};
		
		\path[fill=yellow!20,rounded corners, draw=black!50, dashed]
		(m1) rectangle (m3);           
		\path (vote.north west)+(-1,1) node (m1) {};
		
		\end{pgfonlayer}
		
		\path [draw, ->,very thick] (pol.north)+(-1,1.5) --  node [left] {Polling Control~$(u_k)$} ++(-1,3.3);
		\path [draw, <-,very thick] (pol.north)+(1,1.5) --  node [right] {Observation~$(y_k)$} ++(1,3.3);
		
		\end{tikzpicture}
	\end{scaletikzpicturetowidth}
	\caption{The figure shows a simple hierarchical (influence) network where the individuals are grouped into $N+1$ levels $\text{Level}~0,\text{Level}~1, \cdots, \text{Level}~N$ in a hierarchical fashion.  Each level influences the opinion of the level below it. The network or population has an opinion on the state $x_k$. A pollster samples a subset of individuals and considers the (majority or fraction) opinion $y_k$, runs a local filter to compute the state estimate, and chooses a control to affect the (future) polling mechanism. The aim of the pollster is to estimate the underlying state by controlling the dynamics of polling, while incurring minimum cost.}
	\label{fig:HSN}
\end{figure}
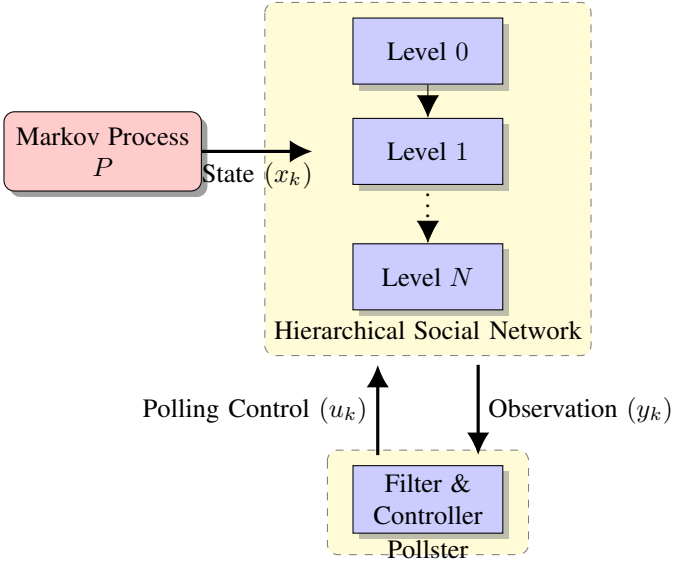 

\subsection{Polling Model and Notation} \label{sec:MN}
Consider the hierarchical social network\footnote{It is to be noted that, the interconnection in the actual social network connecting the people or nodes is irrelevant given the hierarchical influence.} shown in Fig.\ref{fig:HSN}. \\ 
\textbf{State}:~Let $x_k \in \mathcal{X}=~\{ 1, 2, \cdots, X \} $ denote  a Markov chain evolving at discrete time instants $k =0, 1, \cdots $ on a finite state space and observed in noise. This state models the parameters of interest that the population is exposed to. For example, the state $x(\text{Interest~Level}) = \{\text{High},\text{Medium},\text{Low}\}$ could model the interest polling of products; or the state $x(\text{Most~Likely~Candidate}) = \{\text{Cand.1},\text{Cand.2}\cdots\text{Cand.X}\}$ could model electoral polling; etc. \\
\textbf{Pollster's Control/ actions}:~Let $\mathcal{U} = \{1, 2, \cdots, U \}$ denote the set of possible controls (actions), with $u_k \in \mathcal{U}$ denoting the action chosen at time $k$. The action is the decision chosen by the pollster and depends on the polling mechanism (see Sec.\ref{sec:CIP} for adaptive intent polling,  Sec.\ref{sec:CEP} for adaptive expectation polling, and Sec.\ref{sec:CFP} for adaptive friendship polling). \\  Having defined the state and action spaces, we now define the parameters of the POMDP (Sec.\ref{sec:PMDP}). \\
\textbf{Transition matrix}:~Let $P $ denote the homogeneous transition probability matrix of the Markov chain $x_k$ with elements
\begin{equation}
P_{ij} = \mathbb{P}(x_{k+1} = j | x_k = i),~~i,j\in \mathcal{X}.
\end{equation}
Here the transition matrix $P(\neq I)$ captures situations, for example, where political candidates make changes in response to public perception, or the changing interest level in a new app/ product.  \\
\textbf{Polling Cost}: Let $C(x_k,u_k)$ denote the instantaneous cost incurred by the pollster for taking action $u_k$ when in state $x_k$. As will be described in Sec.\ref{sec:CPBO}, this models the measurement cost and quality (accuracy) of the polling mechanism. \\%
\textbf{Opinion}: An opinion indicates the view formed on the underlying state $x_k$ by interacting with other nodes and the environment. Let $\textbf{y}^{l}_{k} \in \mathbb{Y}$ denote the opinion of nodes at level $l$ of the hierarchical network (see Fig.\ref{fig:HSN}). Here $|\mathbb{Y}| = |\mathcal{X}|$ but the elements could be different (see Sec.\ref{ex:Mod}). The observation at different levels in the hierarchical social network assume the following structure: 
observation at the topmost level $\textbf{y}^{0}_{k}$ is directly influenced by the state $x_k$. Observation $\textbf{y}^{l}_{k}$, $l \geq 0$ influences $\textbf{y}^{l+1}_{k}$ (see Fig.\ref{fig:HSN}), i.e,
\begin{equation} \label{eq:Prb_As}
\mathbb{P}(\textbf{y}^{l+1}_{k} = j | \textbf{y}^{l}_{k} = i, x_k = \boldsymbol{x}) \approxeq \mathbb{P}(\textbf{y}^{l+1}_{k} = j | \textbf{y}^{l}_{k} = i).
\end{equation}
\textit{\underline{Discussion of (\ref{eq:Prb_As})}}: The approximation (\ref{eq:Prb_As}) says that the likelihood probabilities of an opinion is mostly influenced by direct superiors or influencers. Even if individuals observe the state, the opinion on the state formed{\footnote{The way individuals form an opinion is not considered in this paper. Behavioural economics models use~(communicative rationality) \cite{HH84}, (motivated reasoning) \cite{Kun90}, to model this.}} depends mostly on more `informed'  sources. The approximation (\ref{eq:Prb_As}) is reasonable in a number of scenarios: knowledge dissemination on the web via Wikipedia, where a minority ($2 \%$) of internet users produce
the content the great majority consumes \cite{Shi10}; in social media like Twitter \cite{BJA11}, where small group of `opinion leaders' gather most of the attention; in judicial hierarchy~\cite{Boy15}, where lower court judges are influenced by their direct superiors and courts above them; etc. \\
The information flow in Fig.\ref{fig:HSN} proceeds according to the following protocol for $k = 0, 1, \cdots$
\begin{compactenum}
	\item The state $x_k$ evolves on time scale~$k$.
	\item Opinions $\textbf{y}^{l}_{k}$ are formed at the Level~$l$ at time $\bar{k} = k + l\delta$. 
	\item At time $k+1$, state transitions to $x_{k+1}$.  
\end{compactenum}
In this protocol, it is assumed that $N\delta \ll 1$. In other words, the state $x_k$ is evolving over a slower time-scale than the time-scale over which the opinions are formed across the network given in Fig.\ref{fig:HSN}. Thus, $\bar{k} = k + l\delta \sim \Time$.\\
%
Let the opinion distribution at Level~$0$ be given by the confusion matrix denoted as $\mathcal{H}_0$, with elements
\begin{equation} \label{eq:Op_d}
(\mathcal{H}_0)_{ij} = \mathbb{P}(\textbf{y}^0_{k} = j | x_{k} = i),~~i\in \mathcal{X}, j\in \mathbb{Y}.
\end{equation}  
The opinions at levels $l \in \{1, \cdots, N\}$ in the hierarchical network are influenced by the preceding levels (see Fig.\ref{fig:HSN}). The confusion matrix $\mathcal{H}^{l}_{l-1}$ between levels $l-1$ and $l$ has elements
\begin{equation} \label{eq:Op_al}
(\mathcal{H}^{l}_{l-1})_{ij} = \mathbb{P}(\textbf{y}^{l}_{k} = j | \textbf{y}^{l-1}_{k} = i),~~\forall i, j~\in \mathbb{Y},~l \in \{1, \cdots, N\}.
\end{equation}
For tractability, assume that the confusion matrix between successive levels is modeled using the same distribution $B$ in (\ref{eq:Op_d}), i.e,~
\begin{equation} \label{eq:Op_dis}
\forall~l \in \{1, \cdots, N\},~\mathcal{H}^{l}_{l-1} = B = \mathcal{H}_0.
\end{equation}
So 
the opinions at levels $l \in \{0,1, \cdots, N\}$ have an effective opinion distribution 
\begin{equation} \label{eq:Obs_lev}
B_{l} = B^{l+1},
\end{equation}
where $B_l$ denotes the opinion distribution at level~$l$. \\
\textbf{Pollster's observation}: Let $\mathcal{Y}$ denote a countable set of observations with $y_k \in \mathcal{Y}$ representing the observations of the underlying state $x_k \in \mathcal{X}$. The observations model the information on the state gathered via views/ sentiments expressed by the nodes or individuals in the hierarchical social network (see Fig.\ref{fig:HSN}). As will be mentioned in Sec.\ref{sec:CPBO}, depending on the polling mechanism, we consider two types of observations for the pollster: \\ (i)~Majority opinion gathering (Sec.\ref{sec:CIP} and Sec.\ref{sec:CEP}), where{\footnote{Here $|\mathcal{X}|$ denotes the cardinality of the set $\mathcal{X}$.}} $|\mathcal{Y}| = |\mathbb{Y}|$. The pollster samples nodes at various levels and considers the majority opinion reported. In case of a tie, an opinion from the majority opinions is randomly selected.\\ (ii)~Fraction opinion gathering (Sec.\ref{sec:CFP}), where{\footnote{Here $S^{Y} = \underbrace{S \times S \times \cdots S}_\text{Y}$ denotes the ususal Cartesian product of sets. }} 
$\mathcal{Y} = \{0,\frac{1}{Y},\frac{2}{Y},\cdots,1\}^{|\mathcal{X}|}$ for some $Y>0$. So in case of fraction opinion gathering, the observations for the pollster are tuples denoting the fraction in favor of each of the states.\\
\textbf{Pollster's Observation distribution}:~Let $\oblvl(u)$ denote the observation probability matrix with elements
\begin{equation} \label{eq:Obs_d}
\oblvl_{ij}(u) = \mathbb{P}(y_{k+1} = j | x_{k+1} = i, u_k = u),~~i\in \mathcal{X}, j\in \mathcal{Y}.
\end{equation}
The observation matrix/ distribution $\oblvl(u)$ models the likelihood of different public opinions given the state, and is different for different polling mechanisms (see Sec.\ref{sec:CPBO}).

%
\subsection{Example illustrating the model:} \label{ex:Mod}
In election polling, the state could be modeled as $x(\text{Likely Candidate}) = \{\text{Cand.1}, \text{Cand.2}\}$, the opinion formed by the nodes at Level~$l$ could be modeled as $\textbf{y}^l \in \mathbb{Y} = \{\text{Consider voting for Cand.1}, \text{Consider voting for Cand.2}\}$. The pollster's observations in case of majority opinion gathering could be modeled as $y \in \mathcal{Y} = \{\text{Cand.1 has majority vote}, \text{Cand.2 has majority vote}\}$ and in case of fraction opinion gathering, the pollster's observations are fractions $y \in \mathcal{Y} = \{\text{Fraction considering Cand.1}\} \times \{\text{Fraction considering Cand.2} \}$. 

\subsection{Partially Observed Markov Decision Process (POMDP)} \label{sec:PMDP}
We now formulate the adaptive polling problem  as a POMDP using the model in Sec.\ref{sec:MN}. We refer to \cite{Kri16} for a detailed treatment of POMDPs - here due to space restrictions we give a very terse description. \\
\textbf{Belief}: Let the probability mass function, termed as the {\em belief}, of the state at time $k-1$ be denoted as 
\begin{equation} \label{eq:PubBel}
\pi_{k-1}(i) = \mathbb{P}(x=i|y_{1},\hdots,y_{k-1})~\text{for}~i \in \mathcal{X}.
\end{equation} 
The state estimate (\ref{eq:PubBel}) is computed from the opinions gathered by the pollster, and is a sufficient statistic~\cite{Kri16} for the history of actions and opinions $\{u_1,y_{1},\hdots,u_{k-1},y_{k-1}\}$. Let the initial estimate be denoted as $\pi_{0} = (\pi_{0}(i),i\in \mathcal{X})$, where $\pi_{0}(i) = \mathbb{P}(x_{0}=i)$. Note that the belief $\pi$ lives in the $X-1$ dimensional unit simplex $\Pi(X) = \{\pi : \pi(i) \in [0,1], \sum_{i=1}^{X} \pi(i) = 1\}.$ We refer to $\Pi(X)$ as the \textit{belief space}. As is well known in POMDPs,  instantaneous cost $C(\pi_k,u_k)$ in terms of the belief $\pi_k$ given by
\begin{equation} \label{eq:Cst_Eq}
C(\pi_k,u_k) = \sum_{i} C(x_k=i,u_k) \pi_k(i),
\end{equation}
where $\pi_k$ denotes the probability distribution function (\ref{eq:PubBel}) at time $k$.
Let~$\rho \in [0,1]$ denote an economic discount factor. An infinite horizon discounted cost POMDP is the tuple $\theta = (\mathcal{X}, \mathcal{Y},\mathcal{U},C,P,\oblvl(u),\rho)$ with  dynamics given by the Bayesian filtering update
\begin{equation} \label{eq:FIL}
\pi_k = T(\pi_{k-1},y_{k},u_{k}),~\text{where}~T(\pi,y,u)= \frac{\oblvl_y(u)P^\prime \pi}{\boldsymbol{1}^\prime \oblvl_y(u)P^\prime \pi}
\end{equation}
and $\oblvl_y(u) = \text{diag}(\oblvl_{1,y}(u), \cdots, \oblvl_{X,y}(u))$. \\
%
Associated with a stationary (time independent) policy $\mu : \Pi(X) \rightarrow \mathcal{U}$ and initial belief $\pi_0 \in \Pi(X)$, is the infinite horizon discounted cost \cite{Kri16}:
\begin{equation} \label{eq:CPC}
J_{\mu}(\pi_0;\theta) = \mathbb{E}_{\mu} \{ \sum_{k=0}^{\infty} \rho^k C(\pi_k,u_k = \mu(\pi_k)) \}.
\end{equation}
Here $J_{\mu}(\pi_0;\theta)$ denotes the cumulative cost for the POMDP model  $\theta$.
The objective of the POMDP is to find the optimal stationary polling policy $\mu^*$ such that
\begin{equation} \label{eq:Opt_J}
J_{\mu^*}(\pi_0;\theta) = {\text{inf}}_{\mu \in \boldsymbol{\mu}} J_{\mu}(\pi_0;\theta)
\end{equation}
where $\boldsymbol{\mu}$ denotes the class of stationary policies. 
Obtaining the optimal policy $\mu^*$ is equivalent to solving Bellman's dynamic programming equation: 
\begin{align} \label{eq:BE}
\mu^*(\pi) &= \argmin_{u \in \mathcal{U}} \mathcal{Q}(\pi,u),~~ J_{\mu^*}(\pi;\theta) = V(\pi), \\
V(\pi) & = \min_{u \in \mathcal{U}} \mathcal{Q}(\pi,u), \nonumber\\
\mathcal{Q}(\pi,u) & = C(\pi,u) + \rho \sum_{y \in \mathcal{Y}} V(T(\pi,y,u)) \sigma(\pi,y,u). \nonumber
\end{align}
Since the belief space $\Pi(X)$ is a continuum, Bellman's equation~(\ref{eq:BE}) does not translate into practical solution methodologies as $V(\pi)$ needs to be evaluated at each $\pi \in \Pi(X)$. The computation of optimal policy of the POMDP is P-SPACE hard \cite{PT87}. The costs $C(\pi,u)$ capture the cost of measurement and the uncertainty or error in the state estimate, and hence are non-linear in the belief; and this results in a non-standard\footnote{POMDP solvers can only handle POMDPs with linear costs, see \cite{Kri16}.} POMDP. This motivates the construction of optimal upper bound policy $\lbpolicy(\pi)$ to the optimal policy $\mu^*(\pi)$ that is inexpensive to compute. In the remainder of the paper, we construct such upper bound policies in terms of easily computable myopic policies for adaptive polling POMDPs.\\
\textit{Remark}: Our adaptive polling formulation in terms of an infinite horizon POMDP (\ref{eq:CPC}) is purely for notational convenience; the optimal policy is stationary. The results in this paper also hold for a finite horizon formulation - then the optimal policy is non-stationary, but Theorem~\ref{thm:MT1} and all subsequent results in this paper continue to hold.

\subsection{Main Result. Optimality of Myopic Polling Policies}
The aim of the pollster is to estimate the (underlying) evolving state~$x_k$ by incurring minimum cumulative cost (\ref{eq:CPC}). The pollster employs the control $u_k = \mu^*(\pi_{k-1})$ to obtain opinions ($y_k \in \mathcal{Y}$) from the nodes, and then updates the belief $\pi_{k-1} \rightarrow \pi_k$~about the underlying state $x_k \in \mathcal{X}$ using~(\ref{eq:FIL}). Theorem~\ref{thm:MT1} below provides sufficient conditions on the observation distribution of the pollster $\oblvl(u)$ such that a myopic polling policy upper bounds the optimal polling policy in (\ref{eq:Opt_J}).\\
Define the myopic policy $\lbpolicy(\pi)$  as
\begin{align}\label{eq:myopiccost}
\lbpolicy(\pi) = \underset{\action \in \actionspace}{\argmin} ~ \cost(\belief,\action)
\end{align} 

\begin{theorem}[Optimality of Myopic Policies via Blackwell Dominance] \label{thm:MT1}
	Consider the adaptive polling POMDP formulated in Sec.\ref{sec:PMDP}. Assume that the cost $C(\pi,u)$ is concave in~$\pi$. Suppose $\oblvl(u) \succeq_B \oblvl(u+1)~\forall u \in \mathcal{U}$. Then  $\lbpolicy(\pi)$ defined in (\ref{eq:myopiccost}) is an upper bound to the optimal polling policy $\mu^*(\pi)$ defined in (\ref{eq:BE}), i.e, $\mu^*(\pi) \leq \lbpolicy(\pi)$ for all $\pi \in \Pi$. In particular, for belief states where $\bar{\mu}(\pi) = 1$, the myopic policy coincides with the optimal policy $\mu^*(\pi)$. 
\end{theorem} 
\textit{\underline{Discussion}}: Theorem~\ref{thm:MT1} is a well known structural result for POMDPs \cite{Kri16}, and it says that if the instantaneous cost of choosing the more accurate/ informative polling control is smaller, then it is the optimal control to choose for the polling POMDP.  \\
(i)~Note that the trivial sub-optimal policy $\hat{\mu}(\pi) = 1 ~\forall \pi \in \Pi(X)$ is also an upper bound to the optimal policy - but an useless upper bound because $\mu^*(\pi) = 1 \implies \hat{\mu}(\pi) = 1$. In comparison, the upper bound constructed via Theorem~\ref{thm:MT1} (Blackwell dominance) says that 
$\bar{\mu}(\pi) = 1\implies \mu^*(\pi) = 1$, which is a much more useful construction. Thus, the myopic polling policy forms a provably optimal upper bound to the computationally intractable optimal policy. \\
(ii)~The usefulness of Theorem~\ref{thm:MT1} stems from the fact that $\lbpolicy(\pi)$ is straightforward  to compute, and the only condition on the model parameters being the concavity of the cost and Blackwell dominance of the observation distributions. \\ Theorem~\ref{thm:MT1} serves as a meta-theorem and sets the stage for the rest of the paper. In the remainder of the paper, we will determine novel sufficient conditions for Blackwell
dominance in the context of adaptive polling. 

\subsection{Blackwell Ordering and R{\'e}nyi Divergence Interpretation} 
R{\'e}nyi Divergence is a generalization of the Kullback-Leibler divergence~\cite{CT02}, and it measures the dissimilarity between two distributions. 
Theorem~\ref{thm:RD1} below shows the relation between R{\'e}nyi Divergence and Blackwell dominance. \\ With a slight abuse of notation in (\ref{eq:Obs_d}), let $\oblvl_i(u)$ denote the $i^{th}$ row of the observation likelihood matrix $\oblvl(u)$. In words, $\oblvl_i(u)$ is the distribution over the observation alphabet $\mathcal{Y}$ conditional on the state $x = i$. \\
\underline{R{\'e}nyi Divergence}: For an observation likelihood $O(u)$, the R{\'e}nyi Divergence of order $\alpha \in [0,1)~\text{for}~ i,j \in \mathcal{X}$ is defined as
\begin{equation} \label{eq:dRD}
D_{\alpha}(\oblvl_i(u)|| \oblvl_j(u)) = \frac{1}{\alpha-1} \log \sum_{y \in \mathcal{Y}} \oblvl^{\alpha}_{iy}(u) \oblvl^{\alpha-1}_{jy}(u).
\end{equation}
\begin{theorem}[Ordering of R{\'e}nyi Divergence] \label{thm:RD1}
	If the observation distribution for the pollster satisfy $\oblvl(u) \succeq_B \oblvl(u+1)~\forall u \in \mathcal{U}$, then for any $ i,j \in \mathcal{X}$:
	\begin{equation}
	D_\alpha(\oblvl_i(u)|| \oblvl_j(u)) \geq D_\alpha(\oblvl_i(u+1)|| \oblvl_j(u+1))~ \forall u \in \mathcal{U}. 
	\end{equation} 
\end{theorem}
\textit{\underline{Discussion}}: Theorem~\ref{thm:RD1} says that when $\oblvl(u) \succeq_B \oblvl(u+1)$, conditional on the state, the observation distributions are more dissimilar in case of $\oblvl(u)$. Here, more the dissimilarity, better the pollster is able to distinguish the states. In Sec.\ref{sec:CPBO}, we discuss the ordering of R{\'e}nyi Divergence for more general channels that arise in hierarchical social networks. Theorem~\ref{thm:RD1} provides a ranking of these general channel structures in the order of their ability to distinguish the states.

\section{Adaptive Polling and Blackwell Dominance} \label{sec:CPBO}
Armed with the POMDP formulation of the previous section, in this section we give sufficient conditions for Blackwell dominance of Theorem~\ref{thm:MT1} to hold for three polling mechanisms:
(i)~Adaptive Intent Polling (Sec.\ref{sec:CIP}), (ii)~Adaptive Expectation Polling (Sec.\ref{sec:CEP}),  (iii)~Adaptive Friendship Polling (Sec.\ref{sec:CFP}). \\
The formulations in this section are adaptive (feedback control based) generalizations of the intent and expectation polling \cite{RW11,DKS12}, and the recently proposed Neighbourhood Expectation Polling \cite{NV18} mechanisms, to account for the time varying state and the hierarchical influence structure present in social networks. 

\subsection{Adaptive Intent Polling} \label{sec:CIP}

In adaptive intent polling, a node at level $l$ is sampled with a probability $\beta_l$ and is asked the following question:
$$``\textit{What does it (a node at level $l$) think the state is?}"$$
This polling mechanism is a more sophisticated version of standard intent polling, for multiple states and hierarchical social networks. In intent polling~\cite{RW11}, to decide between two states, the sampled individuals are asked ``who would you vote for?". \\ 
In the adaptive intent polling formulation below, the pollster adapts the polling policies, namely, the probabilities with which the nodes at different levels in the hierarchical social network are polled. This affects the observation distribution $\oblvl(u)$,  and hence the state estimate~(see Fig.\ref{fig:HSN}).
\subsubsection{Intent Polling Costs}
The instantaneous cost in the adaptive intent polling problem consists of two components-- the measurement cost and the entropy cost (uncertainty in the state estimate):
\begin{compactenum}
	\item[i.)] \textit{\underline{Measurement Cost}}:  Let $u \in \{1,2, \cdots,~U\}$ model the choice of distributions (polling actions) $\beta^{(u)}$, where~$\beta^{(u)} = (\beta^{(u)}_0, \beta^{(u)}_1,\cdots \beta^{(u)}_N)$ and $\sum_i \beta^{(u)}_i = 1$. Here $\beta^{(u)}_l$ for $l = 0,1,2\cdots, N$ denotes the probability of selecting a node from level~$l$, having an opinion distribution $B^{l+1}$.  Let $s(l)$ denote the measurement cost from level $l$. Since nodes at higher levels in the hierarchy (small $l$) provide more informative (in the Blackwell sense) observations, higher costs are associated with obtaining observations from these levels, i.e, $s(l) \geq s(l+1)$, and the average measurement cost for employing the polling mechanism  $\beta^{(u)}$ is $S(\beta^{(u)}) = \sum_{l=0}^{N} \beta^{(u)}_l s(l)$.
	
	\item[ii.)] \textit{\underline{Entropy Cost}}: The entropy cost models the uncertainty in the state estimate~$\pi$~(\ref{eq:PubBel}), and is given as
	$$\eta_E (\pi,u) = - \gamma_1(u) \sum_{i=1}^2 \pi(i) \text{log}_2 \pi(i) + \gamma_2(u) $$
	for $\pi_k(i) \in (0,1)$ and $\eta_E {\overset{\Delta}=} 0$ for $\pi_k(i) = \{ 0,1\}$. Here $\gamma_1, \gamma_2 >0$ are user defined scalar weights. \\ Since more informative opinions lead to larger reduction in uncertainty, $\gamma_1(u) > \gamma_1(u+1)$ and $\gamma_2(u+1) > \gamma_2(u)$.  
	
\end{compactenum}
The instantaneous cost $C(\pi,u)$ in (\ref{eq:Cst_Eq}) incurred by the pollster in case of adaptive intent polling is thus given as:
\begin{equation} \label{eq:CIP}
C(\pi,u) = S(\beta^{(u)}) + \eta_E(\belief,\action).
\end{equation}
The cost (\ref{eq:CIP}) expressed in terms of the belief state $\pi$ captures  the fact  that a control with higher measurement cost should result in a smaller entropy (more reduction in uncertainty) cost and vice versa. 

\subsubsection{Main  Result.  Myopic Policies for Adaptive Intent Polling} \label{subsec:MPBO}
Our main result on adaptive intent polling is Theorem~\ref{cor:CIP} below. It shows that when it cheaper for the pollster to (myopically) listen to the polynomial channel that provides largest reduction in uncertainty on the state, it is indeed optimal to do that. Polynomial channels are parallel cascaded channels that model the communication medium between the pollster and the nodes of a social network having a hierarchical influence structure as in Fig.\ref{fig:HSN}, when the pollster polls all levels of the hierarchical network as in intent polling.\\
%
Let $f_u(z) = \sum_{l=0}^{N} \beta^{(u)}_l z^{l}$ denote the polynomial corresponding to the polling policy $\beta^{(u)}$. For an opinion distribution $B$ (defined in (\ref{eq:Op_dis})), let the matrix polynomials be $f_u(B)~\forall u \in \mathcal{U}$. 
\begin{theorem}[Adaptive Intent Polling] \label{cor:CIP}
	Consider the adaptive intent polling problem with costs specified in (\ref{eq:CIP}). Let the observation distribution for the pollster be{\footnote{The matrix polynomial $f_u(B)$ has an identity observation likelihood for the co-efficient $\beta_0^{(u)}$. This motivates the choice $\oblvl(u) = B f_u(B)~\forall u \in \mathcal{U}$.}} $\oblvl(u) = B f_u(B)~\forall u \in \mathcal{U}$. Assume that the polynomial $f_{U}(z) \in \mathcal{P}_N$ is Hurwitz{\footnote{A polynomial $f$ is Hurwitz if all its zeroes lie in the open left half-plane of the complex plane, and all its co-efficients have the same sign.}}.  \\
	(i)~Then, $\oblvl(u) \succeq_B \oblvl(u+1) ~\forall u \in \mathcal{U}.$\\ 
	(ii)~By Theorem~\ref{thm:MT1}, the myopic intent polling policy $\lbpolicy_I(\pi)$ forms an upper bound to the optimal intent polling policy $\mu_I^*(\pi)$, i.e, $\mu_I^*(\pi) \leq \lbpolicy_I(\pi)$ for all $\pi \in \Pi$. In particular, for belief states where $\bar{\mu}_I(\pi) = 1$, the myopic policy coincides with the optimal policy $\mu_I^*(\pi)$.
\end{theorem}
\textit{\underline{Discussion}}: The instantaneous cost for adaptive intent polling~(\ref{eq:CIP}) is concave in $\pi$ by definition. The proof of Theorem~\ref{cor:CIP} follows from Proposition~\ref{thm:CIP} below and Theorem~\ref{thm:MT1}. The adaptive intent polling mechanism employed by the pollster determines how the opinions are gathered, and the opinions are distributed as $\oblvl(u)$ for the pollster. For an opinion distribution~$B$, the observation distribution of the pollster in case of adaptive intent polling is given as $\oblvl(u) = B f_u(B)$, where $f_u(B) = \sum_{l=0}^{N} \beta^{(u)}_l B^{l}$ and nodes at level $l$ are sampled with probability~$\beta^{(u)}_l $. Proposition~\ref{thm:CIP} below provides a justification for the polynomial $f_{U}(z)$ to be Hurwitz. If $f_{U}(z)$ is Hurwitz, then a way to compute $f_{g}(z)$ for $g \in \{U-1,\cdots,2,1 \}$ is by successive long-division of $f_{U}(z)$ by linear or quadratic factors of $f_{U}(z )$. 
%
%
\subsubsection{ Matrix  polynomials and Blackwell Dominance}
Let~$\mathcal{P}_N = \{ h | h(z) = \sum_{i=0}^N \beta_i z^i,~\sum_{i=0}^N \beta_i = 1,~\beta_i \geq 0 \}$ denote the collection of all polynomials with co-efficients that are a convex combination. 
\begin{proposition} \label{thm:CIP}
	Let $Q$ be a stochastic matrix. 
	For $n>m$, let $p(z) \in \mathcal{P}_n$ and $q(z) \in \mathcal{P}_m$ be two polynomials such that all the roots of $q(z)$ are roots of $p(z)$. If $q(z)$ and $p(z)$ are Hurwitz, then $q(Q) \succeq_B p(Q)$.
\end{proposition}
\textit{\underline{Discussion}}:  
According to Proposition~\ref{thm:CIP}, if the polynomials are Hurwitz and have common factors, a Blackwell dominance relation exists between their corresponding matrix polynomials. If, however, $p(z) \in \mathcal{P}_n$ is not a Hurwitz polynomial, then $q(Q) \succeq_B p(Q)$ only if the polynomial $q(z) \in \mathcal{P}_m$ is the single quadratic factor ($m=2$) corresponding to any conjugate pair of zeros of $p(z)$ having smallest argument in magnitude; see~\cite{BDPW91}.

Proposition~\ref{thm:CIP} provides a way to (partially) order the observation distributions, and hence is useful in choosing a polling action. In adaptive intent polling (Theorem~\ref{cor:CIP}), the degree of the polynomial is the same as the number of levels in the hierarchy (Fig.\ref{fig:HSN}). A polling action in adaptive intent polling corresponds to choosing the (normalized) co-efficients of a polynomial, and these coefficients are the probabilities of polling from the various levels of the social network. From  Proposition~\ref{thm:CIP}, if the two polling actions are such that the corresponding polynomials are Hurwitz and have common factors, then there exists a Blackwell dominance relation between the observation likelihoods corresponding to the two polling actions.\\ 
\underline{Information Theoretic Consequence}: 
Let $I(\mathcal{X};\mathcal{Y}^{(u)})$ denote the mutual information of channel $f_u(B)$ and $\mathcal{C}^{(u)}$ denote the capacity defined in (\ref{eq:SCap}). Let $f^i_u(B)$ denote the $i^{th}$ row of the matrix polynomial $f_u(B)$.
\begin{corollary} \label{cor:CP_S}
	If the channel error probabilities (likelihoods) for the pollster satisfy $f_u(B) \succeq_B f_{u+1}(B) ~\forall u \in \mathcal{U}$, then 
	\begin{compactenum}	
		\item[i.)]	Shannon Capacity Ordering: $\mathcal{C}^{(u)} \geq \mathcal{C}^{(u+1)}~\forall u \in \mathcal{U}$.
		\item[ii.)] R{\'e}nyi Divergence Ordering: $$D_\alpha(f^i_u(B)|| f^j_u(B)) \geq D_\alpha(f^i_{u+1}(B)|| f^j_{u+1}(B))$$ for all $u \in \mathcal{U}$ and for all $i,j \in \mathcal{X}$.	
	\end{compactenum}
\end{corollary}
\textit{\underline{Discussion}}: The proof of Corollary~\ref{cor:CP_S} follows from Theorem~\ref{eq:SCEq} and Theorem~\ref{thm:RD1}. From Corollary~\ref{cor:CP_S}, the Hurwitz polynomial channels are ordered such that the channel that is a sub channel of the other results in a larger reduction in uncertainty on the state.

Together with Proposition~\ref{thm:CIP}, Corollary~\ref{cor:CP_S} provides an interesting link between Hurwitz (stable) polynomials  and channel capacity. From Proposition~\ref{thm:CIP}, those polling actions that result in Hurwitz (stable) polynomials allow decomposition of channels into sub channels that have higher capacity from Corollary~\ref{cor:CP_S}. 


%

\subsection{Adaptive Expectation Polling} \label{sec:CEP}
In adaptive expectation polling, the pollster changes the question instead to 
\begin{align*}
\hspace{-0cm}``\textit{what does a node at level $i$ think}&~\\ 
\textit{the nodes at level $j (<i)$}& \textit{ would report the state as?}"
\end{align*}
This polling mechanism is a more sophisticated version of standard expectation polling, for multiple states and hierarchical social networks. In expectation polling~\cite{RW11}, to decide between two states, the sampled individuals are asked ``who would your friends vote for?". In a hierarchical network, this can be seen as asking ``who would your more influential friends vote for?". \\ In the adaptive expectation polling formulation below, the pollster controls the observation distribution $\oblvl(u)$ by choosing different levels to gather the majority opinion, and this in turn affects the estimate of the state (see Fig.\ref{fig:HSN}).

\subsubsection{Expectation Polling Costs}
The instantaneous cost in the adaptive expectation polling problem consists of two components-- the measurement cost and the uncertainty in the state estimate:
\begin{compactenum}
	\item[i.)] \textit{\underline{Measurement Cost}}:  Let $u \in \{ 1,2, \cdots,~U\}$ model the choice  of levels. In adaptive expectation polling, unlike adaptive intent polling, not all levels are polled. The pollster selects a level~$l$ and asks the nodes at level $l$ to provide information about the other levels. Let $S(u)$ denote the measurement cost for action $u$. Since more informative opinions are costlier to obtain, from Theorem~\ref{cor:CEP}(i) below,~$S(u) \geq S(u+1)~\forall u \in \mathcal{U}$. 
	\item[ii.)] \textit{\underline{State-Estimation error}}: 
	The state-estimation error incurred in choosing action $\action$ is modelled as
	\begin{align}\label{eq:costhistory}
	\begin{aligned}
	\errorh_2(\bar{\state}, \action) = \weight_\action\|\bar{\state} - \belief\|_{2}. 
	\end{aligned}
	\end{align}
	In \eqref{eq:costhistory},  $\belief$ denotes the posterior distribution updated according to $(\ref{eq:FIL})$ and $\bar{\state} \in \{e_1,e_2,\cdots, e_X\}$, where $e_i$ is the unit indicator vector. In \eqref{eq:costhistory} using the law of iterated expectation, $\errorh_2(\belief,\action)$ can be expressed in terms of the belief $\belief$ as follows \cite{Kri16}:
	\begin{align}\label{eq:cost_error}
	\begin{aligned}
	\errorh_2(\belief,\action) = \weight_\action\left(1 - \belief^\prime\belief \right)\\
	\end{aligned}
	\end{align}
	Since more informative opinions lead to smaller state-estimation error, from Theorem~\ref{cor:CEP}(i) below,~$w_{u+1}> w_u$.
\end{compactenum}
The instantaneous cost $C(\pi,u)$ in (\ref{eq:Cst_Eq}) incurred by the pollster in case of adaptive expectation polling is thus given as:
\begin{equation} \label{eq:CEP}
C(\pi,u) = S(u) + \errorh_2(\belief,\action)
\end{equation}
The cost (\ref{eq:CEP}) expressed in terms of the belief state $\pi$ models the fact that asking the nodes at level $i$ to provide information on the opinions of nodes at levels $j (<i)$ is costly, but more informative -- smaller state estimation error. 
%

\subsubsection{Main Result. Myopic policies for Adaptive Expectation Polling}
Our main result in adaptive expectation polling is Theorem~\ref{cor:CEP} below. It shows that when it is cheaper for the pollster to (myopically) listen to the ultrametric{\footnote{A square stochastic matrix $Q$ is \textit{ultrametric} if 
		\begin{enumerate} \label{eq:umm_constraint}
			\item $Q$ is symmetric.
			\item $Q_{ij} \geq \min \{Q_{ik},B_{kj}\}$ for all $i,j,k$.
			\item $Q_{ii} > \max Q_{ik}$ for all $k \neq i$.
\end{enumerate}}} channel that provides the most information on the state, it is indeed optimal to do that. Ultrametric channels are (hidden) cascaded channels that model the communication medium between the pollster and the nodes of a social network having a hierarchical influence structure as in Fig.\ref{fig:HSN}, when the pollster seeks opinions formed at the hidden levels from the levels that are easily accessible.
%
\begin{theorem}[Adaptive Expectation Polling] \label{cor:CEP}
	Consider the adaptive expectation polling problem with costs specified in (\ref{eq:CEP}). Assume that the opinion distribution $B$ (defined in (\ref{eq:Op_dis})) is ultrametric. Let the observation distributions for the pollster be $\oblvl(u) = B_{l}^{l_u/l}~\forall u \in \mathcal{U}$. \\ 
	(i)~For the choice of levels $l_u>l_{u+1}$,  we have $$\oblvl(u) \succeq_B \oblvl(u+1)~\forall u \in \mathcal{U}.$$
	(ii)~By Theorem~\ref{thm:MT1}, the myopic expectation polling policy $\lbpolicy_E(\pi)$ forms an upper bound to the optimal expectation polling policy $\mu^*_E(\pi)$, i.e, $\mu_E^*(\pi) \leq \lbpolicy_E(\pi)$ for all $\pi \in \Pi$. In particular, for belief states where $\bar{\mu}_E(\pi) = 1$, the myopic policy coincides with the optimal policy $\mu_E^*(\pi)$.
\end{theorem}

\textit{\underline{Discussion}}: The instantaneous cost for adaptive expectation polling (\ref{eq:CEP}) is concave in $\pi$ by definition. The proof of Theorem~\ref{cor:CEP} follows from Proposition~\ref{thm:CEP} below and Theorem~\ref{thm:MT1}. The expectation polling mechanism employed by the pollster determines how the opinions are gathered, and the opinions are distributed as $\oblvl(u)$ for the pollster. Proposition~\ref{thm:CEP} below provides a justification for the opinion distribution $B$ to be ultrametric. Note that $B_{l}$ denotes the opinion distribution at level $l$, i.e, $B_{l} = B^{l+1}$ from Fig.\ref{fig:HSN}.  For any $K>0$, clearly $B_{K}^{j+1/K+1} = B_{j}$.  This motivates the choice of the observation distribution of the pollster in case of adaptive expectation polling as $\oblvl(u) = B_{l}^{l_u/l}$, where nodes at level $l$ are polled to provide information of the nodes at level $l_u$. It is easiest (see Sec.\ref{sec:NEx}) to poll nodes at level $N$, so a convenient choice is $\oblvl(u) = B_{N+1}^{l_u/N+1}$.

\subsubsection{Fractional Exponents of Stochastic Matrices and Blackwell Dominance}
For any ultrametric matrix $Q$, the $K^{th}$ root, $Q^{1 / K}$, is also stochastic for any positive integer $K$; see \cite{HL11}. 
\begin{proposition} \label{thm:CEP}
	For any ultrametric matrix $Q$, the following hold for any positive integer $j$:
	\begin{itemize}
		\item[a)] $Q^{j/K} \succeq_B Q^{j}$.
		\item[b)] $Q^{j / K} \succeq_B Q^{(j+1) / K} \hdots \succeq_B Q^{(j+K-1) / K}$
		\item[c)] $Q^{j / (K+1)} \succeq_B Q^{j / (K)}$.
		\item[d)] $Q \succeq_B Q^{j / K}$, for all $j > K$.
	\end{itemize}
\end{proposition}

\textit{\underline{Discussion}}: Clearly, $K^{th}$ integer power of a stochastic matrix is a stochastic matrix. Proposition~\ref{thm:CEP} says that fractional power of certain stochastic matrices, namely ultrametric, are also stochastic. In adaptive expectation polling (Theorem~\ref{cor:CEP}), polling actions correspond to choosing different levels in the hierarchy~(Fig.\ref{fig:HSN}) and soliciting opinions of nodes at other levels. In Proposition~\ref{thm:CEP}, $Q^{j+1/K+1}$  can be used to interpret the notion of node at level $K$ providing information on nodes' opinions at level $j$, and hence provides a way to order the likelihoods corresponding to different polling actions. According to Proposition~\ref{thm:CEP}, when the opinion distribution $B$ in~(\ref{eq:Op_d}) is ultrametric, there exists a Blackwell dominance relation between the observation distributions of the pollster. \\
\underline{Information Theoretic Consequence}: 
Let $I(\mathcal{X};\mathcal{Y}^{(l_u)})$ denote the mutual information of the ultrametric channel $Q^{l_u/K}$ and $\mathcal{C}^{(l_u)}$ denote the capacity defined in (\ref{eq:SCap}). Let $Q^{l_{u} / K}_i$ denotes the $i^{th}$ row of the channel $Q^{l_{u} / K}$.
\begin{corollary} \label{cor:CP_E}
		If the channel error probabilities (likelihoods) for the pollster satisfy $Q^{l_{u} / K} \succeq_B Q^{l_{v} / K}$ for any $K>0$, we have 
	\begin{compactenum}
		\item[i.)] Shannon Capacity Ordering: $ \mathcal{C}^{(l_u)} \geq   \mathcal{C}^{(l_v)}$ for $l_u>l_v$.
		\item[ii.)] R{\'e}nyi Divergence Ordering: $$D_\alpha(Q^{l_{u} / K}_i|| Q^{l_{u} / K}_j) \geq D_\alpha(Q^{l_{v} / K}_i|| Q^{l_{v} / K}_j)$$   for all $u \in \mathcal{U}$ and for all $i,j \in \mathcal{X}$. 
	\end{compactenum}
\end{corollary}
\textit{\underline{Discussion}}: The proof of Corollary~\ref{cor:CP_E} follows from Theorem~\ref{eq:SCEq} and Theorem~\ref{thm:RD1}. Corollary~\ref{cor:CP_E} provides an ordering of R{\'e}nyi Divergence  and Shannon capacity between ultrametric channels $Q^{l_u/K},~K>0,~\forall u \in \mathcal{U}$. From Corollary~\ref{cor:CP_E}, the ultrametric channels are ordered such that the information of nodes at  Level~$0$, for example, revealed by the nodes at Level~$N (\neq 0)$ result in a larger reduction in uncertainty on the state, than opinions from nodes at Level~$N (\neq 0)$. 
\subsection{Adaptive Friendship Polling} \label{sec:CFP}
In this section, the majority reporting is relaxed to obtaining opinion fractions from the nodes. Each polled node gathers the opinion from other nodes at the same level on each state and reports the fraction to the pollster. The question asked by the pollster in case of adaptive friendship polling is 
\begin{align*}
\hspace{-0cm}``\textit{what does a node at level $l$ think the fraction}&~\\ 
\textit{in favor of different states is, at level $l$?}"
\end{align*}
This polling mechanism is a more sophisticated version of Neighborhood Expectation Polling (NEP) \cite{NV18}, for multiple states and hierarchical social networks. NEP is a polling mechanism to decide between two states, and is based on ``\textit{Friendship Paradox}" \cite{Fel91,CR16}. The friendship paradox{\footnote{Friendship Paradox \cite{Fel91}: Let $X$ denote a random node and $Y$ denote a random neighbor. Then $\mathbb{E}d(X) \leq \mathbb{E}d(Y)$, where $d(.)$ denotes the degree.}} is usually expressed as ``\text{on average your friends have more friends than you do}" and the polling based on the friendship paradox is \cite{NV18} ``what is a nodes' estimate of the fraction of votes for a particular candidate?". 
In the adaptive friendship polling formulation below, the pollster controls the observation distribution $\oblvl(u)$ by choosing different levels to gather the information in the form of fractions, and this in turn affects the estimate of the state (see Fig.\ref{fig:HSN}).

\subsubsection{Friendship Polling Costs}
The instantaneous cost in the adaptive friendship polling problem consists of two components-- the measurement cost and the error in the state estimate:
\begin{compactenum}
	\item[i.)] \textit{\underline{Measurement Cost}}:  Let $u \in \{ 1,2, \cdots,~U\}$ denote the control inputs that model the choice  of level. In adaptive friendship polling, the nodes at level $l$ are polled to provide information on the fraction of the nodes at level $l$ in favor of each of the states. Let $S(u)$ denote the measurement cost for action $u$. Since more informative opinion fractions are costlier to obtain, from (\ref{eq:FP_dis}) below, $S(u) \geq S(u+1)~\forall u \in \mathcal{U}$. 
	
	\item[ii.)] \textit{\underline{State-Estimation error}}: 
	The state-estimation error incurred in choosing action $\action$ is modelled as in~(\ref{eq:cost_error})
	\begin{align}\label{eq:cost_error11}
	\begin{aligned}
	\errorh_2(\belief,\action) = \weight_\action\left(1 - \belief^\prime\belief \right)
	\end{aligned}
	\end{align}
	where $w_u> 0$ is a is a user-defined scaling factor. \\ Since more informative opinions lead to smaller state-estimation error, from (\ref{eq:FP_dis}) below, $w_{u+1}> w_u$.
\end{compactenum}
The instantaneous cost $C(\pi,u)$ in (\ref{eq:Cst_Eq}) incurred by the pollster in case of adaptive friendship polling is given as:
\begin{equation} \label{eq:CFP}
C(\pi,u) = S(u) + \errorh_2(\belief,\action)
\end{equation}
\subsubsection{Multinomial distribution and Blackwell Dominance}
%
The adaptive friendship polling mechanism employed by the pollster determines how the opinions are gathered, and the observations for the pollster are tuples reported by the nodes that indicate the fraction in favor of each state. Channels specified by multinomial distributions model the likelihood of opinion counts in favor of different states from different nodes at the same level. 
Let $\nnode \in \{1,2,\cdots,\mathbb{N}\}$ denote the number of nodes accessible (friends with) to nodes at each level in the hierarchical social network. This models the possibility of different individuals or nodes having different friends with $\mathbb{N}$ denoting a finite maximum number. Let the observation alphabet for the pollster be $\mathcal{Y} =  \{ (\frac{\boldsymbol{n}_1}{\nnode},\frac{\boldsymbol{n}_2}{\nnode},\cdots,\frac{\boldsymbol{n}_X}{\nnode})~\forall~\mathcal{N}: \boldsymbol{n}_i \in \mathbb{Z_+},~\sum_i \boldsymbol{n}_i = \nnode\},$  where $\mathbb{Z_+}$ denotes the set of non-negative integers. 
Let $O(l)$ denote the opinion fraction that the pollster receives from level $l$, and has elements
\begin{align} \label{eq:Obs_pol}
(O(l))_{ij} = \mathbb{P}(y^l_{k+1} = j | x_{k+1} = i),&~~
i \in \mathcal{X}, j \in \mathcal{Y}&. \nonumber \\
\text{Here}~j =  (\frac{\boldsymbol{n}^{(j)}_1}{\nnode_j},\frac{\boldsymbol{n}^{(j)}_2}{\nnode_j},\cdots,\frac{\boldsymbol{n}^{(j)}_X}{\nnode_j}),~\nnode_j \in \{1,2,\cdots,\mathbb{N}\},&~\sum_h \boldsymbol{n}^{(j)}_h = \nnode_j. \nonumber\\
\mathbb{P} \Big( y^l_{k+1} = j | x_{k+1} = i \Big) = \frac{\nnode_j!}{\boldsymbol{n}^{(j)}_1! \times \cdots \times \boldsymbol{n}^{(j)}_X!} & \prod_{h=1}^{X} (B_l)_{ih}^{\boldsymbol{n}^{(j)}_h}.
\end{align}
Here  $\nnode_j$ and $\boldsymbol{n}^{(j)}_{i}$ indicate the total and the number in favor of $x = i$ reported and $B_l$ denotes the opinion distribution (\ref{eq:Obs_lev}) at level $l$. The likelihood in (\ref{eq:Obs_pol}) is the well known \textit{multinomial distribution}. \\ 
\textit{Remark}: In case of adaptive friendship polling, the nodes report opinion fractions to the pollster. If instead, the nodes report probabilities with $\mathcal{Y} = [0,1]^{|\mathcal{X}|}$, there is a possibility that the pollster receives biased information. There is a disjunction effect -- the beliefs about the state change when aggregated differently. This is the well known \textit{Simpson's Paradox}; see~\cite{Bly72}. \\
Suppose{\footnote{In Sec.\ref{sec:APB} (specifically,  Algorithm~\ref{algo:appx_blc}) we will see how to obtain (approximate) Blackwell dominance of observation distributions using \textit{Le Cam deficiency} if they are not Blackwell comparable a priori.}} for an opinion distribution $B$ (defined in (\ref{eq:Op_dis})), the opinion fractions corresponding to choosing different levels (polling actions) are ordered as, 
\begin{equation} \label{eq:FP_dis}
\oblvl(u) \succeq_B \oblvl(u+1)~\text{for}~u \in \mathcal{U}.
\end{equation}
It is intuitive that the opinion fractions in (\ref{eq:Obs_pol}) from nodes at level~$i$ are more informative than opinion fractions from nodes at level~$j(>i)$ in Fig.\ref{fig:HSN} owing to obvious Blackwell dominance relation of opinion distributions $B_l$ for $l = i,j$ in~(\ref{eq:Obs_lev}).  
\begin{theorem}[Adaptive Friendship Polling] \label{pro:CFP}
	Consider the adaptive friendship polling problem with costs specified  in (\ref{eq:CFP}). 
	By (\ref{eq:FP_dis}) and Theorem~\ref{thm:MT1}, the myopic friendship polling policy $\lbpolicy_F(\pi)$ forms an upper bound to the optimal friendship polling policy $\mu^*_F(\pi)$, i.e, $\mu_F^*(\pi) \leq \lbpolicy_F(\pi)$ for all $\pi \in \Pi$. In particular, for belief states where $\bar{\mu}_F(\pi) = 1$, the myopic policy coincides with the optimal policy $\mu_F^*(\pi)$.
\end{theorem}
%
\underline{Information Theoretic Consequence}: 
Let $I(\mathcal{X};\mathcal{Y}^{(u)})$ denote the mutual information of channel $\oblvl(u)$ and $\mathcal{C}^{(u)}$ denote the capacity defined in (\ref{eq:SCap}). Let $(\oblvl(u))_{i}$ denote the $i^{th}$ row of the multinomial likelihood $\oblvl(u)$ in  (\ref{eq:Obs_pol}).
\begin{proposition} \label{cor:CP_F}
	For the channel error probabilities (likelihoods) $\oblvl(u) \succeq_B \oblvl(u+1)$, we have 
	\begin{compactenum}
		\item[i.)] Shannon Capacity Ordering: $ \mathcal{C}^{(u)} \geq   \mathcal{C}^{(u+1)}$ for $u \in \mathcal{U}$.
		\item[ii.)] R{\'e}nyi Divergence Ordering: $$D_\alpha((\oblvl(u))_{i}|| (\oblvl(u))_{j}) \geq~D_\alpha((\oblvl(u+1))_{i}|| (\oblvl(u+1)_{j})$$ for all $u \in \mathcal{U}$ and for all $i,j \in \mathcal{X}$.
	\end{compactenum} 	
\end{proposition}
\textit{\underline{Discussion}}: The proof of Proposition~\ref{cor:CP_F} follows from Theorem~\ref{eq:SCEq} and Theorem~\ref{thm:RD1}. When the pollster solicits opinion fractions from the nodes at different levels, the observation likelihood is given by the multinomial (counting) distribution. Proposition~\ref{cor:CP_F} provides an ordering of R{\'e}nyi Divergence and Shannon capacity between channels whose error probabilities are the given by the multinomial distributions of likelihoods $B_l$ in (\ref{eq:Obs_lev}). From Proposition~\ref{cor:CP_F}, the fractions received from Level~$0$ in Fig.\ref{fig:HSN} results in the largest reduction in uncertainty for the pollster.


\section{Approximate Blackwell Dominance, Performance Bounds, and Ordinal Sensitivity} \label{sec:APB}
So far we have discussed sufficient conditions for Blackwell dominance; when these conditions  hold, the optimal adaptive polling policy is provably upper bounded
by a myopic policy.
This section discusses approximate Blackwell dominance and the performance loss due to this approximation.
\subsection{Le Cam  Deficiency} \label{sec:LCD}
Given a collection of matrices, it is important to check whether there exists a Blackwell dominance relation, as Theorem~\ref{thm:MT1} can used to compute inexpensive policies. \textit{What if the pollster would like to choose between different polling mechanisms at each polling epoch to estimate the state?}

Now the control inputs $u \in \mathcal{U}$ correspond to choosing between the different polling mechanisms. 
From discussion in Sec.\ref{sec:CPBO}, it is clear that there is no apparent ordering of observation distributions of the three polling mechanisms. 
In this section, an approximation procedure using \textit{Le Cam deficiency} is provided.\\
\underline{Le Cam deficiency}:
For any two stochastic matrices $W$ and $H$, the \textit{Le Cam deficiency} is 
\begin{equation} \label{eq:lecam}
\delta(W,H) ~~{\overset{\Delta}=} \inf_{R \in \mathcal{M}} \|W - HR \|_\infty,
\end{equation}
where $\mathcal{M}$ denotes the set of all stochastic matrices. The $\inf$ in (\ref{eq:lecam}) is achieved -- this can be shown using \textit{Le Cam randomization} criterion~\cite{SS00}. \textit{Le Cam deficiency} enables to calculate the closest matrix that is Blackwell comparable. 

Numerically, (\ref{eq:lecam})  can be solved as a convex optimization problem using  \textit{CVXOPT} toolbox in Python or \textit{CVX} in Matlab.
Solving (\ref{eq:lecam}) yields observation distributions that are Blackwell comparable. This can be used to obtain Blackwell comparable observation distributions for adaptive friendship polling (Sec.\ref{sec:CFP}). 

%
\begin{algorithm}[t!]
	\caption{Approximate Blackwell Dominance}\label{algo:appx_blc}
	Let $\mathcal{M}$ denotes the set of all stochastic matrices.\\
	\textbf{Initialize}: $\oblvl(1) = \hat{\oblvl}(1)$ \\
	\textbf{For $u \in \{1,2,\cdots,U-1\}$, do}: \\
	$R^*_{u+1} = \argmin_{R \in \mathcal{M}}\|\oblvl(u+1) - \hat{\oblvl}(u) R \|$ \\
	$\hat{\oblvl}(u+1) = \hat{\oblvl}(u)R^*_{u+1}$ \\
	\textbf{end} \\
	\textbf{Output}: $\hat{\oblvl}(u)$ for $u \in \mathcal{U}$.
\end{algorithm}

Consider a POMDP model $\theta = (P,\oblvl(u),C,\rho)$, where $\oblvl(u)$ for $u = \{1,2, \cdots, U\}$ are observation matrices that are not Blackwell comparable. Consider an approximation $\gamma = (P,\oblvl(1),\hat{\oblvl}(\hat{u}),C,\rho)$, where $\hat{u} = \mathcal{U}/\{1\}$ and the observations distributions are such that 
\begin{equation} \label{eq:blc_appx}
\oblvl(1) \succeq_B \hat{\oblvl}(2) \cdots \succeq_B \hat{\oblvl}(U).
\end{equation}
Algorithm~\ref{algo:appx_blc} details a procedure to compute observations distributions that share a Blackwell dominance relation as in~(\ref{eq:blc_appx}).
\subsection{Performance Bounds on Comparison of Polling POMDPs} \label{sec:PB}
Let $\theta = (P,\oblvl(u),C,\rho)$ denote the given adaptive polling POMDP model and $\gamma= (P,\hat{\oblvl}(u),C,\rho)$ denote the adaptive polling POMDP model having a Blackwell dominance relation between the observation distributions. Let $J_{\mu^*(\gamma)}(\pi;\theta)$ and $J_{\mu^*(\gamma)}(\pi;\gamma)$ be defined as in (\ref{eq:CPC}), and denote the cumulative costs incurred by the two models $\theta$ and $\gamma$ respectively, when using the polling policy $\mu^*(\gamma)$.  Let $J_{\mu^*(\theta)}(\pi;\theta)$ and $J_{\mu^*(\theta)}(\pi;\gamma)$ be defined as in (\ref{eq:CPC}), and denote the cumulative costs incurred by the two models $\theta $ and $\gamma$ respectively, when using the polling policy $\mu^*(\theta)$.  Theorem~\ref{res:LCD} below provides a bound on the deviations from the optimal cost and policy performance of the adaptive polling POMDP models.

\begin{theorem} \label{res:LCD}
	Consider two adaptive polling POMDP models $\theta = (P,\oblvl(u))$ and $\gamma = (P,\hat{\oblvl}(u))$ with identical costs and discount factor $(C,\rho)$. Then for the mis-specified model and mis-specified policy, the following sensitivity bounds hold:
	\begin{align} \label{eq:PBCP}
{\hspace{-0.6cm}}	\text{Mis-specified Model:}	&~\sup_{\pi \in \Pi} |J_{\mu^*(\gamma)}(\pi;\gamma) - J_{\mu^*(\gamma)}(\pi;\theta)| \leq G \| \gamma - \theta \|. \\
{\hspace{-0.6cm}}	\text{Mis-specified Policy:}&~J_{\mu^*(\gamma)}(\pi;\theta) \leq J_{\mu^*(\theta)}(\pi;\theta) + 2G  \| \gamma - \theta \|.
	\end{align}
	Here $G = \max_{i \in \mathcal{X},u} \frac{C(e_i,u)}{1-\rho}$ and $e_i$ denotes the indicator vector with a `1' in the $i^{th}$ position, and 
	\begin{equation*} \label{eq:PBCs}
	\| \gamma - \theta \| = \max_{u} \max_i \sum_y \sum_j P_{ij}|\oblvl_{jy}(u) - \hat{\oblvl}_{jy}(u)|.
	\end{equation*}
\end{theorem}
\textit{\underline{Discussion}}:
Theorem~\ref{res:LCD} provides uniform bounds on the additional cost incurred for using parameters that are Blackwell comparable in place of the given parameters of the adaptive polling POMDP. The proof follows from arguments similar to Theorem~$14.9.1$ in \cite{Kri16}, and is omitted. Algorithm~\ref{algo:appx_blc} and Theorem~\ref{res:LCD} can be used to design polling POMDPs that have observation distributions that are not Blackwell comparable, for example, when the polling distributions in case of adaptive intent polling are not Hurwitz; in case of adaptive friendship polling; to name a few.  \\
Proposition~\ref{cor:PB} below highlights the usefulness of Theorem~\ref{res:LCD} in designing polling POMDPs that provide a choice between two polling mechanisms in Sec.\ref{sec:CPBO}.  
Let the opinion distribution  $B$ (defined in (\ref{eq:Op_d})) be ultrametric and $f_2(z) \in \mathcal{P}_N$ be \textit{any} polynomial. Let the true POMDP model be $\theta = (P,\oblvl(1),\oblvl(2),C)$ and the approximation be $\gamma = (P,\oblvl(1),\hat{\oblvl}(2),C)$. Let $\mu(\cdot;\gamma)$ denote the policy parameterized by the approximate model $\gamma$. 
\begin{proposition}[Adaptive Expectation v/s Intent] \label{cor:PB}
	Let $\oblvl(1) = B_{N+1}^{l_1/N+1}$, and $\oblvl(2) = B f_2(B)$  for some $l_1~\text{and}~f_2$, denote the observation distributions in case of  adaptive expectation polling and adaptive intent polling respectively. \\
	(i)~The approximate Blackwell ordering using Algorithm~\ref{algo:appx_blc} is $$\oblvl(1) \succeq_B \hat{\oblvl}(2).$$ 
	(ii)~The myopic polling policy $\lbpolicy(\pi;\gamma)$ is an upper bound to the optimal polling policy $\mu^*(\pi;\gamma)$, i.e, $\mu^*(\pi;\gamma) \leq \lbpolicy(\pi;\gamma)$ for all $\pi \in \Pi$.
\end{proposition}
\textit{\underline{Discussion}}:  For $u=1$, the pollster  chooses expectation polling and hence listens to an ultrametric channel, and for $u=2$, the pollster chooses intent polling and hence listens to a polynomial channel. As $O(2) = Bf_2(B)$, we have $B \succeq_B O(2)$. Note that since $\oblvl(1) = B_{N+1}^{l_1/N+1}$, when $l_1 = 1$ (nodes at Level~$N$ are polled to provide opinion of nodes at Level~$0$), $O(1) = B_{N+1}^{1/N+1} = B \succeq_B O(2)$. This implies that expectation polling is more informative than intent polling.\\ For $l_1>1$, there is no apparent comparison of ultrametric and polynomial channels. However, Algorithm~\ref{algo:appx_blc} can be used to design polling POMDPs for arbitrary $l_1~\text{and}~f_2$. Proposition~\ref{cor:PB} and hence Theorem~\ref{res:LCD} provides the performance bounds for the pollster to choose between the two polling mechanisms.  

\subsection{Ordering of Hierarchical Social Networks} \label{sec:PB_OHSN}

So far we have discussed three types of polling mechanisms on a single hierarchical social network. In this section, we briefly discuss how to order hierarchical networks that differ in the opinion distributions $B$ (defined in (\ref{eq:Op_dis})), according to the expected polling cost. Theorem~\ref{cor:OS_pol} below shows that some networks are inherently more expensive to poll than others; it defines a partial order over networks that results in an ordering of the cost of polling.

Let the POMDP model of the hierarchical network $\mathbb{H}_i$ for $i = 1,2, \cdots$ be $\theta_i$, where the tuple $\theta_i = (P,\oblvl^{(i)},C)$. Let $\mu_i^*(\pi;\theta_i)$ denote the optimal polling policy on each of the network, and let $J_{\mu_i^*(\theta_i)}(\pi;\theta_i)$ denote the corresponding optimal cumulative cost.

\begin{theorem}[Ordinal sensitivity in Polling] \label{cor:OS_pol}
	Consider two hierarchical networks $\mathbb{H}_1$ and $\mathbb{H}_2$. Let the adaptive polling POMDPs for each hierarchical network have the observation distributions that satisfy $\oblvl^{(1)} \succeq_B \oblvl^{(2)}$. Then
	\begin{equation}
	J_{\mu_1^*(\theta_1)}(\pi;\theta_1) \leq J_{\mu_2^*(\theta_2)}(\pi;\theta_2).
	\end{equation}
	Here $\oblvl^{(1)} \succeq_B \oblvl^{(2)}$ denotes $\oblvl^{(1)}(u) \succeq_B \oblvl^{(2)}(u)~\forall~u\in \mathcal{U}$. 
\end{theorem}
\textit{\underline{Discussion}}:  The proof of Theorem~\ref{cor:OS_pol} follows from arguments similar to Theorem~$14.8.1$ in \cite{Kri16}, and is omitted. Since the observation likelihood for the pollster ($\oblvl^{(i)}~\forall i$) depends on the opinion distribution (\ref{eq:Obs_lev}), Theorem~\ref{cor:OS_pol} provides a way to compare the cumulative costs of hierarchical networks with different opinion distributions. The result is useful, in that, a hierarchical network that has more informative opinion distribution at every level compared to another hierarchical network is cheaper to poll on average as the nodes provide more informative opinions.


\section{Numerical Examples} \label{sec:NEx}

The main results of this paper involve using Blackwell dominance to construct  myopic policies that provably  upper bound  the optimal adaptive  polling policy.
In this section, the performance of this myopic upper bound is illustrated using numerical examples for adaptive polling. Let $\Beliefset_1$ represent the set of belief states for which $\cost(\belief,1) < \cost(\belief,\action) ~\forall \action = 2, \cdots, U$. So on the set $\Beliefset_1$, the myopic policy coincides with the optimal policy $\optpolicy(\belief)$. What is the performance loss outside the set  $\Beliefset_1$? Let  $\discountedcost_{\lbpolicy}(\belief_0)$ denote the discounted costs associated with $\lbpolicy(\belief_0)$. We consider the following two measures for measuring the effectiveness of the myopic polling policy: (i)~The  percentage loss in optimality due to using  the myopic policy $\lbpolicy$ instead of  optimal policy $\optpolicy$ is
	\begin{align}\label{eq:percentloss}
	\begin{aligned}
	\epserr_1 = \cfrac{\discountedcost_{\lbpolicy}(\belief_0) - {\discountedcost}_{\optpolicy}(\belief_0)}{{J}_{\optpolicy}(\belief_0)}.
	\end{aligned}
	\end{align}
	In (\ref{eq:percentloss}), the total average cost is evaluated using $1000$ Monte carlo simulations over a horizon of $100$ time units. The optimal cost ${J}_{\optpolicy}(\belief_0)$ is calculated as in (\ref{eq:Opt_J}). Here $\optpolicy$ is the optimal policy of the non-standard (non-linear cost) POMDP, and is solved by brute force\footnote{Software packages available for solving POMDPs require a linear cost. See \text{http://bigbird.comp.nus.edu.sg/pmwiki/farm/appl/} and \text{http://www.pomdp.org/}} using discretization of the belief simplex. The discretization is carried out using Freudenthal triangulation; see \cite{Lov91}.\\
	(ii)~Define the  following discounted cost
	\begin{align}\nonumber
	\begin{aligned}
	\tilde{\discountedcost}_{\optpolicy}(\belief_0) = \E\left\{\sum_{k=1}^{\infty} \discount^{k-1} \tilde{\cost}\left(\belief_\Time, {\optpolicy(\belief_\Time)}\right)\right\},~
	\text{where}, \\ 
	\tilde{\cost}\left(\belief, \optpolicy(\belief)\right) = \begin{cases} {\cost}\left(\belief,1\right)& \belief \in \Beliefset_1\\
	\costm (\belief,U) + \weight_1\errorh_2(\belief,1) & \belief \not\in \Beliefset_1\end{cases}
	\end{aligned}
	\end{align}
	Clearly an upper bound for the percentage loss in optimality due to using the myopic policy $\lbpolicy$ instead of  optimal policy $\optpolicy$ is
	\begin{align}\label{eq:percentloss_L}
	\begin{aligned}
	\epserr_2 = \cfrac{\discountedcost_{\lbpolicy}(\belief_0) - \tilde{\discountedcost}_{\optpolicy}(\belief_0)}{\tilde{\discountedcost}_{\optpolicy}(\belief_0)}.
	\end{aligned}
	\end{align} 
	In (\ref{eq:percentloss_L}), the cumulative discounted cost is evaluated using $1000$ Monte carlo simulations over a horizon of $100$ time units.
\subsection{Example 1: Market Research. Adaptive Expectation Polling with X=3, Y=3, U = 2 and N=1} 
We describe how to estimate the revenue level a movie generates based on the response received on social media platform YouTube. The popularity in the movie is modeled as a $3$ state Markov chain $\Root$  taking values on the state-space $\statespace=\lbrace \text{High}, \text{Medium}, \text{Low} \rbrace $. 
Prior to a movie's release, the production and the media house (proprietor) associated with the movie release a variety of promotional material, in the form of trailer videos, digital billboards, blogs, pre-screenings etc., to advertise the movie. Also,  the critics and those who see the movie before its release will influence the future movie goers by sharing opinions on social media platforms. So production, media house,  and critics are in $\text{Level}~0$ and the common movie goer is $\text{Level}~1$. We will use adaptive expectation polling (Sec.\ref{sec:CEP}) to poll the common movie goer to estimate the performance of a movie, who provide their opinion on YouTube and Twitter. The pollster asks the following question:
\begin{align*}
\hspace{-0cm}``\textit{what does a node at Level $1$ think the nodes at}&~\\ 
\textit{Level~$0~(\text{u=1})$ and Level~$1~(\text{u=2})$ would report}~&\textit{the state as?}"
\end{align*}
In other words, the pollster asks ``what do you think?" and ``what do they think?". So the polling action $u \in \{1,2 \}$ selects the opinion distributions $B_2^{u/2}$. In the example, we estimate $\oblvl$ and $\oblvl^{1/2}$ (see Appendix~\ref{Ap:EM}).

The elements of the transition probability matrix $\tp$ and the opinion probability matrix $\obp_2$ were estimated from the real-data as follows. First, a sample of 30 recent comedy movies were selected. Depending on their box-office revenues, each of these movies were assigned a state from the state-space $\statespace=\lbrace \text{High}, \text{Medium}, \text{Low} \rbrace $. For each of these movies, YouTube comments on their trailers that expressed personal opinions were collected using the Python YouTube API \footnote{\text{https://gdata-python-client.googlecode.com/hg/pydocs/gdata.youtube.html}}. The sentiment associated with each of the comments was identified using the python library - textblob\footnote{\text{http://textblob.readthedocs.org/en/dev/}}. A matrix consisting of number of positive, neutral and negative comments for state of each movie $\lbrace \text{Good}, \text{Neutral}, \text{Bad} \rbrace$ was formed. Using this matrix, the opinion matrix $\obp_2$, given in \eqref{eq:ex_movie} was then obtained by using maximum likelihood estimation algorithm with{\footnote{See Appendix~\ref{Ap:EM}}} ultrametric constraints. This can be used to obtain the opinion distribution $\obp_1$ of Level~$0$. The computed parameters for $\tp$, $\oblvl(1)  = \oblvl^{1/2}(2)$, and $\oblvl(2)$ are as follows:
\begin{align} \label{eq:ex_movie}
\begin{aligned}
\begin{pmatrix}
0.9089  &  0.0281 &   0.0630\\
0.0346  &  0.9433 &   0.0221\\
0.0065  &  0.0138 &   0.9797
\end{pmatrix},
\begin{pmatrix}
0.6382   &   0.1809  &  0.1809\\
0.1809 &   0.6382  &  0.1809\\
0.1809 &   0.1809  &  0.6382
\end{pmatrix}, \\
\begin{pmatrix}
0.4728 &   0.2636 &   0.2636\\
0.2636 &   0.4728 &   0.2636\\
0.2636 &   0.2636 &   0.4728
\end{pmatrix}.
\end{aligned}
\end{align}
For a new (test) movie, depending on which level the observation is obtained from, the pollster updates the probability distribution over the states using the state transition matrix $P$ and the corresponding estimated opinion distribution matrix $\oblvl(1)$ or $\oblvl(2)$.\\
The costs associated with actions $\action = 1, 2$ are chosen as follows:
\begin{align} \label{eq:ex_stock_costs}
\begin{aligned}
S(1) = 0.5, S(2) = 0.25, \weight_1 = 0.5, \weight_2 = 1.
\end{aligned}
\end{align} 
Note that the costs associated with the actions $u=1$ and $u=2$ in \eqref{eq:ex_stock_costs} assume the following structure: $\weight_1 \le \weight_2$ model the accuracy of the observations and $S(1) \geq S(2)$ model the additional cost in expectation polling -- nodes need to be compensated for exhausting their resources gathering information from different levels.  The probabilities in \eqref{eq:ex_movie} and the costs in \eqref{eq:ex_stock_costs} constitute the POMDP parameters of the example discussed in this section. 
The performance of the myopic policy $\lbpolicy(\belief)$ is now evaluated for this POMDP. The percentage loss in optimality $\epserr_1$, given in \eqref{eq:percentloss}, is evaluated by simulation for different values of the discount factor $\discount$ in~Fig.\ref{fig:Tr_ds}. 

\subsection{Example 2: Large Dimensional Example. Adaptive Intent Polling with X=20, Y=20, U = 5 and N = 9} 
\begin{figure}[t!]
	{\hspace{-0.6cm}}\includegraphics[scale=0.3]{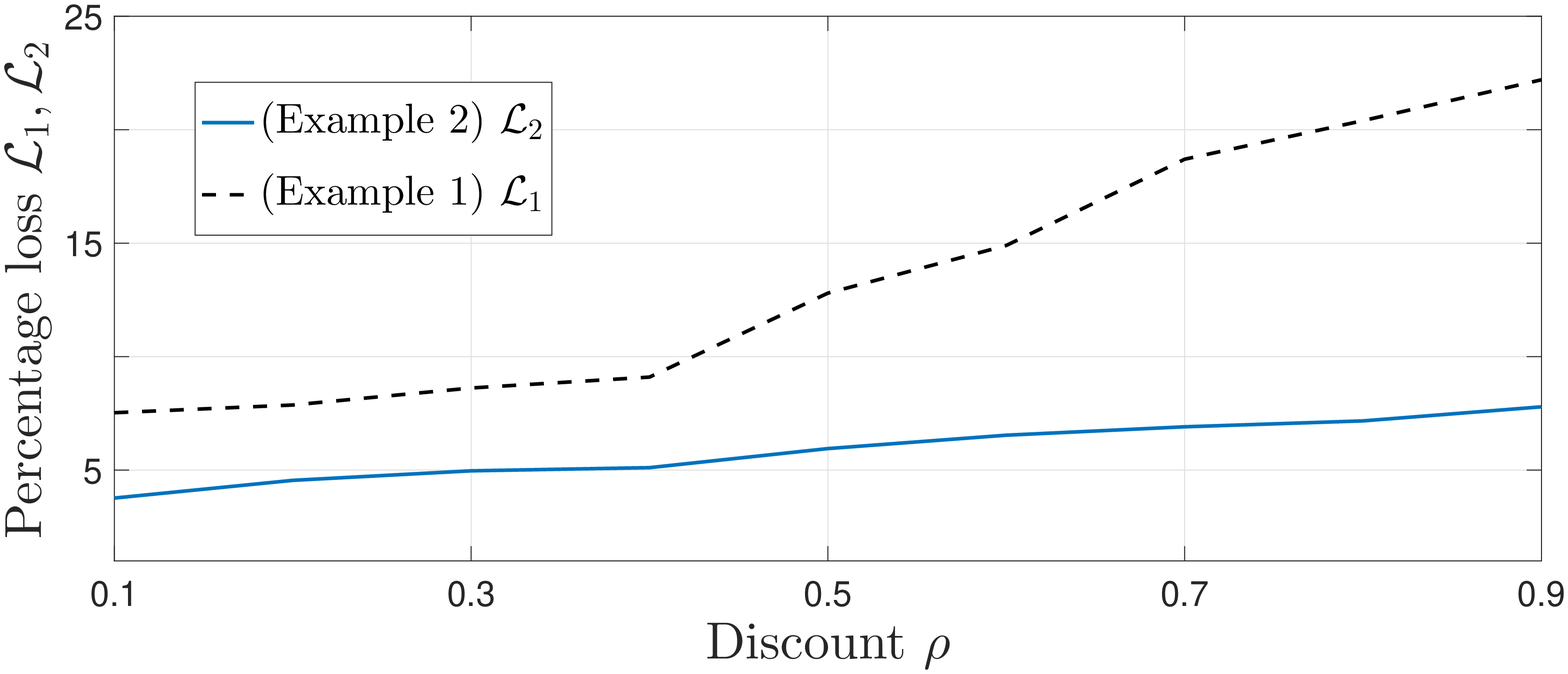}
	\caption{The figure shows the percentage loss in optimality for different values of the discount factor when using a myopic policy.  Here $\epserr_1$ is defined in (\ref{eq:percentloss}) is used as performance loss measure for Example~$1$ and $\epserr_2$ is defined in (\ref{eq:percentloss_L}) is used as a performance loss measure for Example~$2$.  Of course, when $\rho = 0$, the optimal policy is a myopic policy. The percentage loss in Example~$1$ is greater than that in Example~$2$, and this is because of the costs incurred by the optimal policy in (\ref{eq:percentloss}) and (\ref{eq:percentloss_L}).}
	\label{fig:Tr_ds}
\end{figure}

The Blackwell dominance structural result is particularly useful for large number of states and observation symbols since solving the POMDP  (for the optimal policy) is intractable. Random{\footnote{The matrices are generated by stochastic simulation as follows: twenty $(1 \times 20)$ probability vectors were simulated from the Dirichlet distribution on a $19$ dimensional unit simplex and stacked as rows.}} stochastic matrices of size $20 \times 20$ were generated for the transition probability matrix $P$ and the observation probability matrix $B$. We know that $B^{l}$ for $l = 2,\cdots,10$ constitute the opinion distribution of level $l$. The observation distribution of the pollster $O(u) = \sum_{l=0}^{N} \beta^{(u)}_l B^{l+1}$, $\gamma_1 = [5,4,3,2,1]$ and $\gamma_2 = [1,2,3,4,5]$. Here the probability distributions are chosen as{\footnote{$\beta^{(5)} = [25/1296, 1555/15552, 3461/15552, 86925/311040,  13627/62208, \\
		11617/103680, 437/11520, 2671/311040, 73/62208, 29/311040, 1/311040].$} } and $\beta^{(u)}$ for $u = \{4,3,2,1\}$ are obtained by successively removing the smallest root. The (average) percentage loss in optimality $\epserr_2$ calculated using (\ref{eq:percentloss_L}) is shown in the in Fig.\ref{fig:Tr_ds}. The percentage loss in Fig.\ref{fig:Tr_ds} is calculated as the average of $\epserr_2$ over  $10$ pairs of random $20 \times 20$ stochastic matrices for $P$ and $B$.

\section{Conclusions}
This paper considered the problem of adaptive (feedback control based) polling in hierarchical social networks, and the problem was formulated as a partially observed Markov decision process (POMDP). We presented three main results. First, we presented an adaptive generalization of intent polling to hierarchical social networks. The notion of Blackwell dominance was extended to the case of polynomial observation likelihoods (channels) described by matrix polynomials. Second, we presented an adaptive generalization of expectation polling to hierarchical social networks. The notion of Blackwell dominance was extended to the case of ultrametric observation likelihoods (channels) described by fractional matrix powers. Third, we presented an adaptive generalization of neighborhood expectation polling to hierarchical social networks. The notion of Blackwell dominance was extended to the case of multinomial distributions of observation likelihoods.

This extension of Blackwell dominance to more general channels that arise in hierarchical social networks was used to provide a natural ordering of  R{\'e}nyi Divergence and Shannon capacity. These information theoretic consequences provide a ranking of these general channel structures in the order of their ability to distinguish the states.  \\
We also discussed approximate Blackwell dominance based on Le Cam deficiency to facilitate the comparison of the different polling mechanisms, and situations where a Blackwell dominance relation is absent. Performance bounds on the cumulative cost and polling policy were provided when the model parameters are mis-specified. \\
Finally, the results and the performance of the myopic polling policy was illustrated on a dataset from YouTube. 

\appendices
\section{Proofs}

\textbf{\underline{Proof of Theorem~\ref{thm:MT1}}}: \\
Denote by $\Obs^{(\action)}$ as the observations recorded when using action $\action$. Then  $\oblvl(u+1) = \oblvl(u) R$ implies the following
\begin{align}\label{eq:relationBnew}
\prob\left(\Obs^{(\action+1)}|\Root\right) = \sum_{\Obs^{(\action)}}\prob\left(\Obs^{(\action+1)}|\Obs^{(\action)}\right)\prob\left(\Obs^{(\action)}|\Root\right)
\end{align}

For notational convenience, let $\filter(\belief,\obs,\action)$ be written as $\filter(\belief,\Obs^{(\action)}= \obs)$. Observe that,
\begin{align}\label{eq:proof}
\begin{aligned}
\filter\left(\belief,\Obs^{(\action+1)}=\obs\right) =& \cfrac{\oblvl_{\action+1}(\obs)\tp^\prime\belief}{\filternorm\left(\belief,\Obs^{(\action+1)} = \obs\right)}
=\sum_{r}\pmfnew(r)\filter(\belief,\Obs^{(\action)} = r)
\end{aligned}
\end{align}
where $\pmfnew(r)$ is a probability mass function w.r.t $r$ and defined as
\begin{align} \label{eq:pmfnew}
\pmfnew(r) = \prob\left(\Obs^{(\action+1)}= \obs|\Obs^{(\action)} = r\right)\cfrac{\filternorm\left(\belief,\Obs^{(\action)} = r\right)}{\filternorm\left(\belief,\Obs^{(\action+1)} = y\right)}
\end{align}
The following inequality follows from the concavity of $\valuefunction(\belief)$  and \eqref{eq:pmfnew}
\begin{align}\label{eq:impl}
\begin{aligned}
\valuefunction\left(\filter\left(\belief,\Obs^{(\action+1)}=\obs\right) \right) = & \valuefunction\left(\sum_{r}\pmfnew(r)\filter(\belief,\Obs^{(\action)} = r)\right)\\
\valuefunction\left(\filter\left(\belief,\Obs^{(\action+1)}=\obs\right) \right) \geq & \sum_{r}\pmfnew(r)\valuefunction\left(\filter(\belief,\Obs^{(\action)} = r)\right)\\
\end{aligned}
\end{align}


Following completes the proof of Theorem \ref{thm:MT1} using \eqref{eq:impl}.
\begin{multline}
\sum_{\obs}\filternorm(\belief, \Obs^{(\action+1)} = \obs)   \valuefunction\left(\filter\left(\belief,\Obs^{(\action+1)}=\obs\right) \right) \geq  \\
\sum_{\obs}\sum_{r}\pmfnew(r)\valuefunction  \left(\filter(\belief,\Obs^{(\action)} = r)\right)\filternorm(\belief, \Obs^{(\action+1)} = \obs)\\
= \sum_{r}\valuefunction\left(\filter\left(\belief,\Obs^{(\action)}=r\right)\right)  {\filternorm\left(\belief,\Obs^{(\action)} = r\right)}
\end{multline}
$\therefore  \cost(\belief,1) \le \cost(\belief,\action) ~\forall \action \Rightarrow \optpolicy(\belief) = 1 \Rightarrow \optpolicy(\belief) \le \lbpolicy(\belief). \qed$
\vspace{0.0cm}

\textbf{\underline{Proof of Theorem~\ref{thm:RD1}}}: \\
Let $\oblvl(u) \succeq_B \oblvl(u+1)$ for $u \in \mathcal{U}$. From the definition of R{\'e}nyi Divergence (\ref{eq:dRD}) we have \cite{NJ12}:
\begin{multline} \label{eq:DrD}
{\hspace{-0.5cm}}D_\alpha(\oblvl_i(u+1)|| \oblvl_j(u+1)) \leq  \min \Big\{  (1-\alpha) D(\oblvl_i(u+1)|| \oblvl_j(u+1)), \\ \alpha D(\oblvl_j(u+1)|| \oblvl_i(u+1)) \Big\}.
\end{multline} 
We know that \cite{Sak64}:
\begin{multline} \label{eq:SaI}
{\vspace{-0cm}}\oblvl(u) \succeq_B \oblvl(u+1) \Rightarrow \\ D(\oblvl_i(u)|| \oblvl_j(u)) \geq D(\oblvl_i(u+1)|| \oblvl_j(u+1)),
\end{multline} 
for all $i,j \in \mathcal{X}$. From (\ref{eq:DrD}) and (\ref{eq:SaI}), the result follows. \qed
\vspace{0.0cm}

\textbf{\underline{Proof of Theorem~\ref{thm:CIP}}}: \\
It is given that $p(z) \in \mathcal{P}_n$ and $q(z) \in \mathcal{P}_m$, with $n>m$. Clearly, $f(Q)$ and $g(Q)$ are stochastic matrices. Further, if the quotient polynomial $h(z) = \frac{f(z)}{g(z)} \in \mathcal{P}_{(n-m)}$, then it is easily seen that $g(Q) \succeq_B f(Q)$. \\  Since the polynomials $p(z)$ and $q(z)$ are Hurwitz, the quotient polynomial $h(z) = \frac{p(z)}{q(z)} = \sum_{i=0}^{(n-m)} \alpha_i z^i$ has positive co-efficients; i.e $\alpha_i >0$. It suffices to prove that $h(z)\in \mathcal{P}_{(n-m)}$. It is clear that $p(1) = q(1) = 1$, which implies that $h(1) = 1$; i.e,~$\sum_{i=0}^{(n-m)} \alpha_i = 1$. \qed 
\vspace{0.0cm}

\textbf{\underline{Proof of Theorem~\ref{thm:CEP}}}: \\
We will only prove Theorem~\ref{thm:CEP}b and Theorem~\ref{thm:CEP}c. \\
For Theorem~\ref{thm:CEP}b, we have  $Q^{(j+J)/K} = Q^{j/K} \times Q^{J/K} $. Therefore $Q^{j/K} \succeq_B Q^{(j+J)/K}$.\\
For Theorem~\ref{thm:CEP}c, we have $Q^{j/K} = Q^{j/K+1} \times Q^{j/K(K+1)} $. Therefore $Q^{j/K} \succeq_B Q^{j/K+1}$. \qed 
\section{EM Algorithm with Ultrametric Constraints} \label{Ap:EM}
The parameters of the POMDP are computed using a sequence of observations obtained from level ${\actiondim}$ in Fig.\ref{fig:HSN}. Specifically, a modified version of the EM algorithm~\cite{DLR77} is used to compute the maximum likelihood estimate of the tuple $\left(\tp, \obp_{\actiondim+1}\right)$, where $\obp_{\actiondim+1}$ is restricted to the space of ultrametric stochastic matrices. The opinion probability matrices at all other levels are computed by taking fractional exponents of $\obp_{\actiondim+1}$. In the modified EM algorithm, computing $\obp_{\actiondim+1}$ requires maximizing an auxiliary likelihood function (of observation sequences) subject to ultrametric constraints (see Footnote~$13$) on $\obp_{\actiondim+1}$. However, the space of ultrametric stochastic matrices is non-convex because of constraint $\obp_{\actiondim+1}(i,j) \ge \min\left\{\obp_{\actiondim+1}(i,k), \obp_{\actiondim+1}(k,j) \right\}$ and thus computationally intractable. \\ The following reformulation based on the Big-M method in linear programming \cite{GNS09} is used to deal with the non-convex constraint.  For all $ i,j,k \in \statespace, i \neq j \neq k$:
\begin{align}\label{eq:cvx_relaxation}
\obp_{\actiondim+1}(i,j) &\geq \obp_{\actiondim+1}(i,k) + M (1 - \kappa), \\
\obp_{\actiondim+1}(i,j) &\geq \obp_{\actiondim+1}(k,j) + M  \kappa, \\
\obp_{\actiondim+1}(k,j) & \geq \obp_{\actiondim+1}(i,k) + M (1 - \kappa), \\
\obp_{\actiondim+1}(i,k) & \geq \obp_{\actiondim+1}(k,j) +M  \kappa, \\
\kappa & \geq 0, \\
-\kappa & \geq -1,  
\end{align}
for some large positive value~$M$. The resulting observation likelihood $B_{N+1}$ is a stochastic and ultrametric matrix.


\begin{thebibliography}{10}
\providecommand{\url}[1]{#1}
\csname url@samestyle\endcsname
\providecommand{\newblock}{\relax}
\providecommand{\bibinfo}[2]{#2}
\providecommand{\BIBentrySTDinterwordspacing}{\spaceskip=0pt\relax}
\providecommand{\BIBentryALTinterwordstretchfactor}{4}
\providecommand{\BIBentryALTinterwordspacing}{\spaceskip=\fontdimen2\font plus
\BIBentryALTinterwordstretchfactor\fontdimen3\font minus
  \fontdimen4\font\relax}
\providecommand{\BIBforeignlanguage}[2]{{%
\expandafter\ifx\csname l@#1\endcsname\relax
\typeout{** WARNING: IEEEtran.bst: No hyphenation pattern has been}%
\typeout{** loaded for the language `#1'. Using the pattern for}%
\typeout{** the default language instead.}%
\else
\language=\csname l@#1\endcsname
\fi
#2}}
\providecommand{\BIBdecl}{\relax}
\BIBdecl

\bibitem{Lec12}
L.~Le~Cam, \emph{Asymptotic methods in statistical decision theory}.\hskip 1em
  plus 0.5em minus 0.4em\relax Springer Science \& Business Media, 2012.

\bibitem{Lec12a}
L.~Le~Cam and G.~L. Yang, \emph{Asymptotics in statistics: some basic
  concepts}.\hskip 1em plus 0.5em minus 0.4em\relax Springer Science \&
  Business Media, 2012.

\bibitem{Ber16}
D.~Bergemann and S.~Morris, ``Bayes correlated equilibrium and the comparison
  of information structures in games,'' \emph{Theoretical Economics}, vol.~11,
  no.~2, pp. 487--522, 2016.

\bibitem{Rei91}
U.~Rieder, ``Structural results for partially observed control models,''
  \emph{Methods and Models of Operations Research}, vol.~35, no.~6, pp.
  473--490, 1991.

\bibitem{Kri16}
V.~Krishnamurthy, \emph{Partially Observed Markov Decision Processes}.\hskip
  1em plus 0.5em minus 0.4em\relax Cambridge University Press, 2016.

\bibitem{CKZ98}
J.~Cohen, J.~Kempermann, and G.~Zbaganu, \emph{Comparisons of Stochastic
  Matrices with Applications in Information Theory, Statistics, Economics and
  Population}.\hskip 1em plus 0.5em minus 0.4em\relax Springer Science \&
  Business Media, 1998.

\bibitem{LMS02}
L.~L{\'o}pez, J.~F. Mendes, and M.~A. Sanju{\'a}n, ``Hierarchical social
  networks and information flow,'' \emph{Physica A: Statistical Mechanics and
  its Applications}, vol. 316, no.~1, pp. 695--708, 2002.

\bibitem{ALS03}
J.~A. Almendral, L.~L{\'o}pez, and M.~A. Sanju{\'a}n, ``Information flow in
  generalized hierarchical networks,'' \emph{Physica A: Statistical Mechanics
  and its Applications}, vol. 324, no.~1, pp. 424--429, 2003.

\bibitem{Shi10}
C.~Shirky, \emph{Cognitive surplus: Creativity and generosity in a connected
  age}.\hskip 1em plus 0.5em minus 0.4em\relax Penguin UK, 2010.

\bibitem{BJA11}
J.~Borge-Holthoefer, A.~Rivero, I.~Garc{\'\i}a, E.~Cauh{\'e}, A.~Ferrer,
  D.~Ferrer, D.~Francos, D.~Iniguez, M.~P. P{\'e}rez, G.~Ruiz \emph{et~al.},
  ``Structural and dynamical patterns on online social networks: the spanish
  may 15th movement as a case study,'' \emph{PloS one}, vol.~6, no.~8, p.
  e23883, 2011.

\bibitem{Boy15}
C.~L. Boyd, ``The hierarchical influence of courts of appeals on district
  courts,'' \emph{The Journal of Legal Studies}, vol.~44, no.~1, pp. 113--141,
  2015.

\bibitem{Lov91a}
W.~S. Lovejoy, ``A survey of algorithmic methods for partially observed markov
  decision processes,'' \emph{Annals of Operations Research}, vol.~28, no.~1,
  pp. 47--65, 1991.

\bibitem{PT87}
C.~H. Papadimitriou and J.~N. Tsitsiklis, ``{The complexity of Markov decision
  processes},'' \emph{Mathematics of operations research}, vol.~12, no.~3, pp.
  441--450, 1987.

\bibitem{Bla53}
D.~Blackwell, ``Equivalent comparisons of experiments,'' \emph{The Annals of
  Mathematical Statistics}, pp. 265--272, 1953.

\bibitem{CT02}
T.~M. Cover and J.~A. Thomas, \emph{Elements of information theory}.\hskip 1em
  plus 0.5em minus 0.4em\relax John Wiley \& Sons, 2012.

\bibitem{Rag11}
M.~Raginsky, ``{Shannon meets Blackwell and Le Cam: Channels, codes, and
  statistical experiments},'' in \emph{Information Theory Proceedings (ISIT),
  2011 IEEE International Symposium on}.\hskip 1em plus 0.5em minus 0.4em\relax
  IEEE, 2011, pp. 1220--1224.

\bibitem{RBOJBW17}
J.~Rauh, P.~K. Banerjee, E.~Olbrich, J.~Jost, N.~Bertschinger, and D.~Wolpert,
  ``{Coarse-graining and the Blackwell order},'' \emph{Entropy}, vol.~19,
  no.~10, p. 527, 2017.

\bibitem{RW11}
D.~M. Rothschild and J.~Wolfers, ``{Forecasting elections: Voter intentions
  versus expectations},'' 2011.

\bibitem{NV18}
B.~Nettasinghe and V.~Krishnamurthy, ``{What Do Your Friends Think? Efficient
  Polling Methods for Networks Using Friendship Paradox},'' \emph{arXiv
  preprint arXiv:1802.06505}, 2018.

\bibitem{Fel91}
S.~L. Feld, ``Why your friends have more friends than you do,'' \emph{American
  Journal of Sociology}, vol.~96, no.~6, pp. 1464--1477, 1991.

\bibitem{KH14}
V.~Krishnamurthy and W.~Hoiles, ``Online {R}eputation and {P}olling {S}ystems:
  {D}ata {I}ncest, {S}ocial {L}earning, and {R}evealed {P}references,''
  \emph{{IEEE} Trans. Comput. Social Systems}, vol.~1, no.~3, pp. 164--179,
  2014.

\bibitem{DKS12}
A.~Dasgupta, R.~Kumar, and D.~Sivakumar, ``Social sampling,'' in
  \emph{Proceedings of the 18th ACM SIGKDD international conference on
  Knowledge discovery and data mining}.\hskip 1em plus 0.5em minus 0.4em\relax
  ACM, 2012, pp. 235--243.

\bibitem{Sat94}
E.~Satake, ``{Bayesian Inference in Polling Technique: 1992 Presidential
  Polls},'' \emph{Communication Research}, vol.~21, no.~3, pp. 396--407, 1994.

\bibitem{Lin13}
D.~A. Linzer, ``{Dynamic Bayesian forecasting of presidential elections in the
  states},'' \emph{Journal of the American Statistical Association}, vol. 108,
  no. 501, pp. 124--134, 2013.

\bibitem{HH84}
J.~Habermas and J.~Habermas, \emph{The theory of communicative action}.\hskip
  1em plus 0.5em minus 0.4em\relax Beacon press, 1984, vol.~2.

\bibitem{Kun90}
Z.~Kunda, ``The case for motivated reasoning.'' \emph{Psychological bulletin},
  vol. 108, no.~3, p. 480, 1990.

\bibitem{BDPW91}
R.~Barnard, W.~Dayawansa, K.~Pearce, and D.~Weinberg, ``Polynomials with
  nonnegative coefficients,'' \emph{Proceedings of the American Mathematical
  Society}, vol. 113, no.~1, pp. 77--85, 1991.

\bibitem{HL11}
N.~J. Higham and L.~Lin, ``On pth roots of stochastic matrices,'' \emph{Linear
  Algebra and its Applications}, vol. 435, no.~3, pp. 448--463, 2011.

\bibitem{CR16}
Y.~Cao and S.~Ross, ``The friendship paradox,'' \emph{Mathematical Scientist},
  vol.~41, no.~1, 2016.

\bibitem{Bly72}
C.~R. Blyth, ``On simpson's paradox and the sure-thing principle,''
  \emph{Journal of the American Statistical Association}, vol.~67, no. 338, pp.
  364--366, 1972.

\bibitem{SS00}
V.~Spokoiny and A.~Shiryayev, \emph{{Statistical Experiments And Decision,
  Asymptotic Theory}}.\hskip 1em plus 0.5em minus 0.4em\relax World Scientific,
  2000, vol.~8.

\bibitem{Lov91}
W.~S. Lovejoy, ``Computationally feasible bounds for partially observed
  {M}arkov decision processes,'' \emph{Operations research}, vol.~39, no.~1,
  pp. 162--175, 1991.

\bibitem{NJ12}
M.~Naghshvar and T.~Javidi, ``{Active hypothesis testing: Sequentiality and
  adaptivity gains},'' in \emph{46th Annual Conference on Information Sciences
  and Systems (CISS)}.\hskip 1em plus 0.5em minus 0.4em\relax IEEE, 2012, pp.
  1--6.

\bibitem{Sak64}
M.~Sakaguchi, \emph{Information theory and decision making}.\hskip 1em plus
  0.5em minus 0.4em\relax Statistics Dept., George Washington University, 1964.

\bibitem{DLR77}
A.~P. Dempster, N.~M. Laird, and D.~B. Rubin, ``Maximum likelihood from
  incomplete data via the em algorithm,'' \emph{Journal of the Royal
  Statistical Society. Series B (methodological)}, pp. 1--38, 1977.

\bibitem{GNS09}
I.~Griva, S.~G. Nash, and A.~Sofer, \emph{Linear and {N}onlinear
  optimization}.\hskip 1em plus 0.5em minus 0.4em\relax Siam, 2009, vol. 108.


\end{thebibliography}
\end{document}